\documentclass[10pt,prb,aps,twocolumn,showpacs,floatfix]{revtex4}

\usepackage{graphicx}
\usepackage{amssymb}
\usepackage{subfigure}
\usepackage{color}
\usepackage{amsmath} 
\newcommand{\half}{\hbox{$\frac{1}{2}$}}

\begin{document}

\title{Finite-size scaling and boundary effects in two-dimensional valence-bond-solids}

\author{Anders W. Sandvik}
\affiliation{Department of Physics, Boston University, 590 Commonwealth Avenue, Boston, Massachusetts 02215, USA}

\begin{abstract}
Various lattice geometries and boundary conditions are used to investigate valence-bond-solid (VBS) ordering in the ground state of an 
$S=1/2$ square-lattice quantum spin model---the $J$-$Q$ model, in which four- or six-spin interactions $Q$ are added to the standard 
Heisenberg exchange $J$. Ground state results for finite systems (with up to thousands of spins) are obtained using an unbiased projector
quantum Monte Carlo method. It is found that great care has to be taken when extrapolating the order parameter to infinite lattice size, 
in particular in cylinder geometry. Even though strong VBS order exists in two dimensions, and is established clearly with increasing 
system size on $L\times L$ lattices (or $L_x\times L_y$ lattices with a fixed aspect ratio $L_x/L_y$ of order $1$), only short-range VBS 
correlations are observed on long cylinders (when $L_x \to \infty$ at fixed $L_y$). The correlation length increases with the cylinder width, 
until long-range order sets in at a ``critical'' width. This width is very large even when the 2D order is relatively strong. For example, 
for a system in which the order parameter is $70\%$ of the largest possible value, $L_y=8$ is required for ordering. Extrapolations of the 
VBS order parameter based on correlation functions (the square of the order parameter) for small  $L \times L$ lattices can also be misleading. 
For a $20\%$-ordered system results for $L$ up to $\approx 20$ appear to extrapolate clearly to a vanishing order parameter, while for larger 
lattices the scaling behavior crosses over and extrapolates to a non-zero value (with exponentially small finite size corrections). The VBS order 
parameter also exhibits interesting edge effects related the known emergent U($1$) symmetry close to a ``deconfined'' critical point, which, if not 
considered properly, can lead to wrong conclusions for the thermodynamic limit. The observed finite-size behavior for small $L\times L$ lattices 
and long cylinders is very similar to that predicted for a Z$_2$ spin liquid. The results therefore raise concerns about recent numerical work 
claiming Z$_2$ spin liquid ground states in 2D frustrated quantum spin systems, in particular, the Heisenberg model with nearest and 
next-nearest-neighbor couplings. Based on the results presented so far, a VBS state in this system cannot be ruled out.

\end{abstract}

\date{\today}


\maketitle

\section{Introduction}
\label{sec:intro}

A valence-bond solid (VBS) is a state of a quantum spin system in which there is no magnetic long-range order, but lattice symmetries 
(translational and some times rotational) are broken due to dimerization or, more generally, polymerization of the system into one with 
a larger unit cell than the underlying lattice. One can think of the spins within a unit cell of a VBS (or within different groups of spins 
in a large complex unit cell) as having an enhanced probability of forming a total spin singlet. In this paper, manifestations of VBS order 
in ground states of finite systems are investigated, using unbiased quantum Monte Carlo (QMC) simulations of $S=1/2$ spins on the 
two-dimensional (2D) square lattice with interactions---Heisenberg exchange supplemented by certain multi-spin interactions---leading 
to columnar order in the thermodynamic limit.\cite{sandvik07} The approach to the infinite-size 2D limit is investigated for different boundary 
conditions. The models considered can be tuned from strong to weak VBS order (and also through a critical point), enabling bench-mark investigations 
of asymptotics and cross-over behaviors. In particular, consequences of near-criticality of the VBS order on the finite-size behavior can be examined 
in detail. The stability of VBS order on long cylinders ($L_x\times L_y$ lattices with $L_x \gg L_y$) is also addressed. This geometry is often used 
in density matrix renormalization (DMRG) studies,\cite{white07,stoudenmire12} with recent intriguing results pointing to the absence of VBS order and the 
existence of spin liquids in frustrated models whose ground states have been debated for a long time.\cite{jiang08,yan11,jiang11} 

In the following introductory sections, several background facts motivating further studies of VBS order are discussed and some of the known 
properties of VBS states are briefly reviewed. The purposes of the studies reported here will then be detailed, followed by an outline of rest of the paper.

\subsection{VBS states and frustrated interactions}
\label{sec:frustvbs}

VBS states have been known for a long time to exist in 1D frustrated quantum spin chains. In particular, in the $S=1/2$ Heisenberg 
chain with nearest- and next-nearest-neighbor couplings $J_1$ and $J_2$, the ground state at coupling ratio $g=J_2/J_1=1/2$ is exactly a product of 
singlets formed on alternating nearest-neighbor bonds (a pattern which can be realized in two different ways; hence the ground state is two-fold 
degenerate).\cite{majumdar69} Away from this special, exactly solvable point, there are fluctuations modifying the simple product state. Numerical 
exact diagonalization studies have shown that long-range dimerization survives down to $g_c \approx 0.241$.\cite{nomura92,eggert96,sandvik10a} 
For $g < g_c$ the ground state exhibits critical spin and VBS correlations (like the standard Heisenberg chain with $J_2=0$).\cite{affleck85}  
At higher $g$, the simple dimer VBS order persists at least up to $g \approx 0.6$, above which more complicated VBS or spiral spin states 
likely form.\cite{bursill95,kumar10}

\begin{figure}
\includegraphics[width=5.5cm, clip]{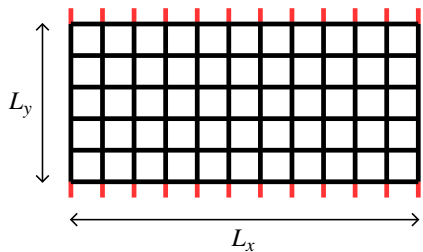}
\vskip-2mm
\caption{(Color online) A cylindrical, semi-periodic 2D square lattice with open edges (left and right sides) and periodic boundary conditions 
in the vertical direction (i.e., the open links at the top and bottom are connected to each other).}
\label{clatt}
\vskip-2mm
\end{figure}

The frustrated 2D square-lattice $J_1$-$J_2$ Heisenberg model (with nearest-neighbor couplings $J_1$ and the $J_2$-interactions connecting spins across the diagonals 
of each four-spin plaquette) also has a non-magnetic ground state in some window of coupling ratios $0.4 \alt g \alt 0.6$ (outside of which the ground state 
is N\'eel antiferromagnetic for smaller $g$ and exhibits stripe antiferromagnetic order for larger $g$).\cite{chandra88,dagotto89,rea89,singh90} However, in this 
case it has been difficult to determine the exact nature of the ground state. Many studies over the past two decades have suggested a VBS, with either columnar 
or plaquette (four-spin unit cell) order,\cite{dagotto89,rea89,singh90,read91,schulz96,kotov99,singh99,sushkov01,capriotti00,mambrini06,beach09,murg09,isaev09} 
but spin liquid ground states (which have no broken symmetries but may have topological order \cite{wen03}) have also been proposed.\cite{chandra88,capriotti01} 
Very recently, results of DMRG calculations on cylindrical semi-periodic lattices (with open edges in one direction---see Fig.~\ref{clatt}) were used to argue more 
specifically that the ground state of the system for $0.41 \le g \le 0.62$ is a $Z_2$ spin liquid.\cite{jiang11} A concurrent calculation based on tensor-product 
states also claimed the absence of VBS order.\cite{wang11} 

A story similar to that of the $J_1$-$J_2$ Heisenberg model has played out in recent years in the case of the $S=1/2$ Heisenberg model with only
nearest-neighbor interactions on the geometrically frustrated kagome lattice. Many calculations initially suggested a VBS ground state (in this case 
with a complex 12- or 36-site unit cell),\cite{marston91,nicolic03,singh07,evenbly10,poilblanc10} but the most recent DMRG studies support a $Z_2$ spin 
liquid scenario \cite{jiang08,yan11} (as also predicted in early analytical work \cite{sachdev92}). Here, as well, cylindrical lattices played 
a crucial role in obtaining the numerical data.

\subsection{Deconfined quantum-critical points and VBSs with emergent U(1) symmetry}
\label{sub:dqc}

When the N\'eel order of a 2D antiferromagnet such as the $S=1/2$ Heisenberg model is destroyed in a continuous quantum phase transition, one scenario is that the putative 
spin-liquid state is immediately unstable to the formation of a VBS. This has been argued to lead to a ``deconfined'' quantum-critical point separating the N\'eel
and VBS states.\cite{senthil04a,senthil04b} The phase transition is associated with deconfinement of spinons. Being generically continuous, due to subtle quantum
interference effectes, this type of transition violates the classical ``Landau rule'', according to which a transition between two ordered states breaking unrelated 
symmetries should be generically first-order. 

In the low-energy field-theory argued to describe the deconfined quantum-critical point (the 2+1 dimensional non-compact CP$^1$ theory),\cite{senthil04a} 
the VBS fluctuations correspond to a U($1$) gauge field to which spinons are coupled. There is a dangerously irrelevant operator 
(a quadrupled monopole operator) which reduces the U($1$) symmetry to a four-fold ($Z_4$) symmetry inside the ordered VBS state 
(in which the spinons become confined). On the square lattice, this corresponds to the four degenerate columnar VBS patterns. Close to the critical point, 
the Z$_4$ symmetry only becomes apparent beyond a length-scale $\Lambda$, which is larger than the standard correlation length $\xi$ associated with the magnitude 
of the order parameter. At distances below $\Lambda$ there are angular fluctuations of the VBS order parameter $(D_x,D_y)$, which in a system with $x$ order,
$(|D_x|>0,D_y=0)$, induces $D_y$ order on length scales up to $\Lambda$, with this length diverging as $\Lambda \sim \xi^{1+a}$ with $a>0$. At distances much below
$\Lambda$, the angle of the VBS order parameter fluctuates in an essentially U($1$) isotropic manner.

The deconfinement scenario appears to be realized in a class of ``$J$-$Q$'' models,\cite{sandvik07,melko08,lou09,sandvik10,sandvik11,banerjee11b} in which the Heisenberg 
exchange $J$ is supplemented by certain multi-spin interactions---products of two or more two-spin singlet projectors acting on different spin pairs. These interactions 
lead to the formation of local correlated singlets, thereby reducing, and eventually destroying, the N\'eel order. Results of QMC calculations (which are not affected by sign problems 
in this case) are consistent with a single critical point separating the N\'eel state and a VBS. While some works suggested that the transition is weakly 
first-order, \cite{jiang08,kuklov08} the most recent studies point to a continuous transition with anomalously large scaling corrections.\cite{sandvik10,sandvik11,banerjee11b} 
Moreover, emergent U($1)$ symmetry has been explicitly observed in the VBS order parameter distribution.\cite{sandvik07,jiang08} By studying the U($1$)--Z$_4$ cross-over, 
the exponent $a$ was estimated in Ref.~\onlinecite{lou09} to be $a = 0.20 \pm 0.05$. 

\subsection{Stability of the spin liquid}

For the frustrated spin systems discussed above in Sec.~\ref{sec:frustvbs}, deconfined quantum-criticality, i.e., a gapless spin liquid existing only at a singular point, 
is also an alternative to the transition out of the N\'eel state into an extended spin liquid phase. At the heart of this issue is the question of the stability of the 
spin liquid state.\cite{hermele04} The deconfined quantum-criticality scenario implies that some spin liquids are generically unstable, at least under some commonly
satisfied conditions, but stable spin liquids can also exist. 

Recently Cano and Fendley succeeded in constructing a long-sought local (but complicated) Hamiltonian \cite{cano10} that is the parent Hamiltonian of the prototypical 
resonating valence-bond (RVB) spin liquid, i.e., the equal superposition of all nearest-neighbor valence bond configurations (with the Marshall sign rule built 
in).\cite{anderson87,sutherland88,liang88} This state, however, is a U($1)$ spin liquid with exponentially decaying spin correlations but critical VBS 
correlations,\cite{tang11b,albuquerque10} and not the kind of fully gapped Z$_2$ spin liquid proposed in the context of the frustrated models discussed above [but a U($1$) 
spin liquid is also a possible ground state candidate of this model\cite{iqbal11}]. As a consequence of its close relationship with the critical Rokhasr-Kivelson dimer 
model,\cite{tang11b,albuquerque10} one would expect this state to be generically unstable to perturbations of the Cano-Fendley Hamiltonian, leading to the formation of 
a VBS. Viewed from the perspective of a class of quantum states, the introduction of longer bonds either maintains the critical VBS,\cite{tang11b} or leads to a Z$_2$ 
spin liquid,\cite{sandvik06,yao11} but the hamiltonian for these extended RVB states is not known.

Stable $Z_2$ spin liquids are known with Klein Hamiltonians on particular decorated lattices,\cite{raman05} but the degree if stability of these states when moving 
away from the limit of high decoration is not known. The Kitaev honeycomb-lattice model,\cite{kitaev06a} which has a $Z_2$ liquid state, can also be related to a model 
of SU($2$) interacting spins on a decorated honeycomb lattice.\cite{fawang10} However, there is still no rigorously known example of a Z$_2$ spin liquid ground state of 
a local SU($2$) invariant Hamiltonian on one of the simple standard 2D lattices (square, triangular, honeycomb, kagome, etc). This lack of a prototypical system underlies the quest to 
find $Z_2$ liquids in numerical studies of frustrated quantum spin Hamiltonians.\cite{capriotti01,jiang11,wang11,jiang08,yan11} Z$_2$ spin liquid states have already 
been confirmed in QMC studies of frustrated quantum XY models.\cite{dang11}

\subsection{Detection of spin liquids and VBS order}

It is highly non-trivial to unambiguously confirm 2D spin liquid states based on numerical calculations on relatively small lattices. The main 
difficulty here is to exclude weak VBS order (while the absence of magnetic order is easier to confirm, e.g., by demonstrating a non-zero spin gap). 
There is therefore much interest in finding positive signals for various spin liquid phases, e.g., using unique finite-size scaling properties of the 
entanglement entropy.\cite{kitaev06,levin06} Other signals related to the topological aspects of spin liquids have also been proposed.\cite{yao11,jiang11}
However, regardless of what properties are investigated, great care has to be taken in view of the small lattices accessible for systems with frustrated 
interactions. Due to sign problems, unbiased QMC studies of the ground states of these systems are essentially impossible \cite{henelius00} (although some 
progress has been made here recently at elevated temperatures \cite{nyfeler08}). Variational QMC methods can be used \cite{capriotti01,clark11} but are not reliable, 
because very different states can have almost the same energy. Exact diagonalization studies can reach $\approx 42$ spins,\cite{richter10,nakano11,lauchli11} 
while DMRG calculations now can reach hundreds of spins.\cite{stoudenmire12} Tensor-product state methods (which can be regarded as generalizations of the 
matrix-product based \cite{schollwock11} DMRG scheme) can reach much larger sizes, but are complicated by the fact that extrapolations also have to be carried 
out in the bond dimension of the tensors.\cite{bauer09,zhao10,wang11} In DMRG calculations there is a similar issue with regards to the maximum number of states 
that can be kept, which is what limits the accessible system sizes (since that number of states in this case has to grow exponentially with the system size). 

\subsection{Lattice shapes and boundaries}

As already mentioned, in DMRG studies it has become popular to use lattices in the form of cylinders with semi-periodic boundary conditions (with periodic 
boundaries along the long direction and open short edges), as illustrated in Fig.~\ref{clatt}. An aspect ratio $L_x/L_y>1$ improves the convergence with the 
number of states kept, as compared to a fully periodic lattice with equal length in both directions (for a given total number of lattices sites).\cite{white07}  
The better convergence with samples of this shape can be traced to the inherently 1D nature of the DMRG procedures and how the generated states can incorporate 
entanglement.\cite{schollwock11} It has also been argued that cylindrical $L_x/L_y>1$ samples, some times in combination with modifications of the 
boundaries  (e.g., using field terms breaking some symmetry), have other favorable effects as well on the convergence of various order parameters as a 
function of the system size.\cite{white07,stoudenmire12} 

In QMC studies of sign-problem free models, periodic $L\times L$ lattices are normally used. In cases where the couplings are spatially anisotropic, 
it has proved helpful to use $L_x\times L_y$ lattices with $L_x \not= L_y$,\cite{sandvik99,jiang12} while in other cases no particular advantages of such 
rectangular lattices were noted.\cite{sandvik10a} Open boundaries have been considered in QMC work primarily in cases where the perturbing effects of the 
edges are the actual targets of investigation.\cite{hoglund09,kaul08} In a previous QMC study of a VBS state it was also noted that open boundaries can be 
used to break the four-fold symmetry of the 2D VBS completely and stabilize a unique VBS pattern, as an alternative of studying VBS correlation functions in 
periodic lattices with no explicitly broken symmetries.\cite{sandvik02}

\subsection{Purpose of the paper}

The main purpose of the present paper is to systematically investigate the role of the lattice shape and boundary conditions on the finite-size scaling 
properties of the VBS order parameter. VBS states have in the past few years been conclusively demonstrated in several 2D $J$-$Q$ 
models,\cite{sandvik07,melko08,jiang08b,lou09,banerjee11a} and also in 1D chains (where the same kind of dimerization transition takes place as in 
the frustrated $J_1$-$J_2$ chain)\cite{tang11,sanyal11} and 3D systems.\cite{beach07} Different types of VBS patterns can be realized, depending on the 
arrangements of the singlet projectors on the lattice. These models have been studied with large-scale QMC simulations, mainly for the purpose of investigating the 
nature of the N\'eel--VBS transition.\cite{sandvik07,melko08,jiang08b,lou09,banerjee11a,sandvik10,sandvik11,banerjee11b} Here the main focus will instead 
be on the VBS state itself (including its cross-over behavior close to criticality), using the $J$-$Q$ models to obtain generic bench-marks for finite-size scaling 
of this kind of order parameter. An efficient approximation-free ground state projector QMC method \cite{sandvik05,sandvik10b} was used to obtain results for both 
strongly and weakly VBS ordered systems on square lattices with different shapes and boundaries. 

In order to make contact with the currently favored manner of applying the DMRG method,\cite{white07,stoudenmire12} cylindrical systems with open edges in one 
direction  will be studied extensively. The convention adopted here is that the edges parallel to the $y$-axis are open, and periodic boundary conditions are applied 
in the other direction. Such an $L_x \times L_y$ lattice is illustrated in Fig.~\ref{clatt}. In some cases the open edges will be modified to favor a certain VBS 
pattern, which is often also done in DMRG studies.\cite{yan11,jiang11} Fully periodic lattices will also be considered. Two aspect ratios, $L_x/L_y=1,2$, will be 
considered for both the semiperiodic and fully periodic systems. The limit $L_x \to \infty$ will also be taken for small $L_y$. 

In addition to suggesting optimal approaches for extracting the VBS order in the 2D thermodynamic limit, the results presented here will also show that the issue 
of excluding VBS order in a system with an unknown type of non-magnetic ground state may be more difficult than what has been anticipated so far. In particular, 
the geometry of long cylinders can give misleading results. Not only can calculations on such systems completely miss 2D VBS order (because the system is
disordered with a short correlation length on the cylinders), but also the claimed positive signals of a 2D Z$_2$ spin liquid \cite{yao11,jiang11} cannot be 
trusted when used with cylinders of practically accessible widths (because they are essentially 1D spin liquids although the state orders in the
2D limit). The emergent U($1$) symmetry of the VBS state leads to interesting boundary effects, which are also studied here.

\subsection{Outline of the paper}

In Sec.~\ref{sec:models} the $J$-$Q$ models are specified in detail, the correlation functions of interest are defined, the projector QMC method is briefly 
outlined, and its convergence properties are discussed and illustrated with an example. Extrapolations of the infinite-size value of the order parameter is 
discussed in Sec.~\ref{sec:extrap}. Results for the $J$-$Q_3$ model at $J=0$ (the pure $Q_3$ model), which has very robust columnar VBS order, is discussed 
first, in order to show how the different ways of extrapolating the order parameter to the thermodynamic limit agree well with each other. Results for three 
different lattice types are compared; periodic $L \times L$ and $2L \times L$ systems as well as semi-periodic cylindrical $2L\times L$ systems. The much weaker 
VBS ordering in the $Q_2$ and  $J$-$Q_2$ models is discussed next, using the same lattices as above. Here several subtle issues are pointed out that affect extrapolations
to infinite size when the order is not strong, and, therefore, the length-scales $\xi$ and $\Lambda$ are large. The quantum-critical scaling form of the VBS order 
parameter is also discussed, as a nearby critical point also influences the finite-size behavior in systems off criticality. The vector aspects of the columnar 
VBS order parameter $(D_x,D_y)$ and the effects of its emergent U($1$) symmetry are studied in detail in Sec.~\ref{sec:boundary}. The evolution of the $x$- and 
$y$-components of the order parameter as a function of the distance from an open edge is studied, with and without symmetry-breaking modifications of the edge. 
In Sec.~\ref{sec:cylinder} the destruction of VBS order on cylinders is studied in the limit $L_x \to \infty$ and $L_y$ fixed. The most important results are 
summarized and their implications are discussed in Sec.~\ref{sec:summary}. Here detailed comparisons with the recent DMRG results \cite{jiang11} for the $J_1$-$J_2$ 
Heisenberg model are also made. Detection of the U($1$)--Z$_4$ symmetry of the VBS order parameter based on probability distributions $P(D_x,D_y)$ generated 
in QMC calculations is discussed in Appendix \ref{appa}.

\section{Models and methods}
\label{sec:models}

\subsection{J-Q models}

\begin{figure}
\includegraphics[width=4.5cm, clip]{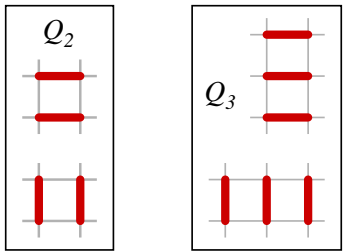}
\vskip-1mm
\caption{(Color online) $Q_2$ and $Q_3$ terms on the square lattice. The bars of length one lattice constant indicate
the locations of singlet projectors $C(i,j)$ on site pairs $i,j$. The Hamiltonian contains all unique translations of 
these operators.}
\label{qterms}
\vskip-1mm
\end{figure}

A generic $J$-$Q$ model is defined using products of singlet projectors $C(i,j)$ on two sites,
\begin{equation}
C(i,j)=\hbox{$\frac{1}{4}$}-{\bf S}_i \cdot {\bf S}_j.
\label{sproj}
\end{equation}
The standard Heisenberg model is just a sum of such singlet projectors over the interacting bonds  $\langle i,j\rangle$ (here nearest-neighbor sites 
on the square lattice),
\begin{equation}
H_J = -J\sum_{\langle i,j\rangle} C(i,j),
\label{hj}
\end{equation}
where the minus sign corresponds to antiferromagnetic interactions. A $Q_n$ term consists of products of two or more ($n$) singlet
projectors acting on different bonds;
\begin{equation}
H_{Q_{n}}=-Q_n\sum_{a}\prod_{b=1}^n C(i[a,b],j[a,b]).
\label{hq}
\end{equation}
Here $a$ is a label corresponding to the lattice units within which the singlet projectors are arranged and $b$ labels the bonds (spin pairs) on which
the singlet projectors within these units act; $i[a,b]$ and $j[a,b]$ above refer to the two sites connected by bond $b$ in unit $a$. In the simplest kind of 
$Q_2$ term on the square lattice, $a$ denotes $2\times 2$ plaquettes, with the two projectors within these plaquettes connecting spins either horizontally or 
vertically (i.e., for a given $2\times 2$ plaquette there are two labels $a$; one corresponding to horizontal and one to vertical bonds). This standard $Q_2$ 
term will be considered here, along with a similar $Q_3$ term with the projectors arranged in columns. Both these cases are illustrated in Fig.~\ref{qterms}. 
In general, the sum over projectors is such that the Hamiltonian does not break any of the symmetries of the lattice. 

The $J$-$Q_n$ model defined by the Hamiltonian $H=H_J+H_{Q_n}$ hosts a VBS ground state when $Q_n/J$ is sufficiently large. In general, VBS formation is favored for 
a large enough number $n$ of singlet projectors (with the minimum being typically $n=2$ or $n=3$ in two dimensions) if the arrangement of them is compatible with 
some symmetry-breaking pattern of strong and weak bond singlets. In this paper the pure $Q_2$ and $Q_3$ models without any $J$ term will be studied primarily, 
but some results for $J$-$Q_2$ systems with $J/Q_2>0$ will also be presented.

\subsection{Projector QMC}

$J$-$Q$ models with minus signs as in Eqs.~(\ref{hj}) and (\ref{hq}) do not have QMC sign problems and can be studied with very
efficient QMC loop algorithms. Here the ground state projector method developed in Ref.~\onlinecite{sandvik10b} is used. It is based
on applying a high power of the Hamiltonian to a ``trial'' state $|\Psi_0\rangle$,
\begin{equation}
|\Psi_m\rangle = (-H)^m |\Psi_0\rangle,
\label{hmproj}
\end{equation}
where $(-H)^m$ is written as a sum over all possible strings of the individual $J$ and $Q$ terms in (\ref{hj}) and (\ref{hq}). Denoting such 
a string of singlet projectors by $P_m(i)$, with $i$ formally indexing the different strings, an operator expectation value is written as
\begin{equation}
\langle A \rangle_m = \frac{\sum_{ij}\langle \Psi_0| P^T_m(j) A P_m(i) | \Psi_0\rangle}{\sum_{ij}\langle \Psi_0 | P^T_m(j) P_m(i) | \Psi_0\rangle},
\label{expval}
\end{equation}
where $P^T_m(j)$ is the string $P_m(j)$ in reverse order. 

The QMC method implements importance sampling of the operator strings $P^T_m(j)P_m(i)$, which is done in two steps, 
as described in detail in Ref.~\onlinecite{sandvik10b} in the case of the Heisenberg model (and the modifications of the scheme when a $Q$ term is 
present are straight-forward and have been discussed briefly in Ref.~\onlinecite{sandvik10a}): First all the operators in the $J$ and $Q$ terms are split 
into their diagonal and off-diagonal components in the basis of spin states $|S^z_1,\ldots,S^z_N\rangle$ used. The diagonal operators can be moved around 
on the lattice as long as each operator is compatible with the spin state on which it acts (with only operations on anti-parallel spins allowed). The 
full set of operators is sampled by changing the types of some operators from diagonal to off-diagonal, or vise versa, on the same lattice unit $a$, 
using an efficient loop algorithm.\cite{evertz93,evertz03,sandvik99b}). 

The ground state of a bipartite $J$-$Q$ model (i.e., with each singlet projector connecting two spins on different sublattices) being guaranteed to be singlet, it is
particularly convenient to use a trial state expressed in the valence bond basis in the singlet sector. The convergence of $\langle A \rangle_m$ to the true ground 
state expectation value $\langle 0| A|0 \rangle_m$ is then dictated by the gap to the second singlet. For a periodic lattice (or a semi-periodic cylinder), a 
transitional-invariant trial state also filters out excited states with non-zero momentum from the outset. Translational invariance in the applicable 
lattice direction(s) is easily ensured by using an amplitude-product state \cite{liang88} for $|\Psi_0\rangle$, i.e., a superposition written in terms of 
bipartite valence bond states $|v\rangle$,
\begin{equation}
|\Psi_0\rangle = \sum_v c_v |v\rangle.
\end{equation}
Here the sum includes all tilings of the $N$-site lattice into $N/2$ bipartite two-spin singlets, i.e.,
\begin{equation}
|v\rangle = \left | \prod_{i=1}^{N/2} (i,j_i^v) \right \rangle,
\end{equation}
where $(i,j)=(|\uparrow_i\downarrow_j\rangle - |\downarrow_i\uparrow_j\rangle)/\sqrt{2}$ with $i$ and $j$ sites on sublattice $A$ and $B$, respectively,
and the weight $c_v$ of a given tiling $v$ into singlets depends only on the ``shapes'' ${\bf l}=(l_x,l_y)$ of the bonds in $|v\rangle$;
\begin{equation}
c_v = \prod_{\bf l} h({\bf l})^{n_{\bf l}},
\end{equation}
where $n_{\bf l}$ is the number of bonds of shape ${\bf l}$. 

Amplitude-product states are very easy to sample in the course of the projection according to (\ref{hmproj}), as also described in Ref.~\onlinecite{sandvik10b}. 
The detailed form of the amplitude $h({\bf l})$ is not crucial when the state is used as a trial state. Variationally optimized amplitudes lead to faster convergence 
with the power $m$, but even without optimizing the convergence properties are good.\cite{sandvik10b} In the work reported here, amplitudes decaying with the bond length 
$l$ as $l^{-3}$ were used (in which case the trial state itself has N\'eel order, but this is very quickly destroyed by the projection procedure
in a VBS state).

\subsection{Correlation functions}

In order to characterize the ground state, the spin ($s$) and dimer ($d$) correlation functions are computed. These are defined in the standard way as
\begin{eqnarray}
C_s ({\bf r}_{ij}) & = & \bigl \langle {\bf S}({\bf r}_i) \cdot {\bf S}({\bf r}_j) \bigr \rangle, \label{csdef} \\
C_{d\alpha} ({\bf r}_{ij}) & = & \bigl \langle B_\alpha({\bf r}_i)B_\alpha({\bf r}_j) \bigr \rangle, \label{cddef}
\end{eqnarray}
where ${\bf r}_{ij}={\bf r}_{i}-{\bf r}_{j}$ is the spatial separation of the operators and $B_\alpha$, $\alpha=\hat x,\hat y$, 
is the dimer operator on nearest-neighbor bonds oriented in the $\alpha$ direction, e.g., for $\alpha=\hat x$
\begin{equation}
B_{\hat x}({\bf r})={\bf S}({\bf r}) \cdot {\bf S}({\bf r}+\hat {\bf x}).
\label{bxdef}
\end{equation}
One can also define cross-correlations $\langle B_{\hat x}({\bf r}_i)B_{\hat y}({\bf r}_j) \rangle$ but they will not be needed here. The correlation
functions can be easily computed using loop estimators based on the transition graphs generated when the sampled valence bond states in the ket and bra 
states of Eq.~(\ref{expval}) are propagated by the string of singlet projectors. The estimators are discussed in detail in Refs.~\onlinecite{tang11} 
and \onlinecite{beach06}.

Columnar and plaquette VBS states can both be detected by the columnar VBS order parameter, which when averaged over the whole
lattice of $N=L_xL_y$ sites can be defined by the operators
\begin{eqnarray}
D_x & = & \frac{1}{N}\sum_{x,y}B_{\hat x}(x,y)(-1)^x, \label{dxsum} \\
D_y & = & \frac{1}{N}\sum_{x,y}B_{\hat y}(x,y)(-1)^y. \label{dysum}
\end{eqnarray}
In a columnar state with the lattice rotational symmetry completely broken, either $D_x$ or $D_y$ has a non-zero expectation value, while
in a plaquette state they are both non-zero and equal. The $J$-$Q$ models studied here host only columnar VBS states. However, as we will be
discussed below, in a columnar state on a finite lattice one can have both non-zero $\langle D_x\rangle$ and $\langle D_y\rangle$, due 
to boundary and shape effects.

In periodic and semi-periodic systems where the degeneracy of the possible VBS patterns is not broken, one can only detect the VBS with the
corresponding correlation functions, e.g., the squares of the order parameters defined above. In particular, it is useful to consider the
total squared order parameter,
\begin{eqnarray}
D^2 = D_x^2 + D_y^2.
\label{d2def}
\end{eqnarray}
The magnitude of the order parameter in a corresponding symmetry-broken state is $D=\langle D^2\rangle^{1/2}$ (which can be taken as a definition of the value 
$D$ of the order parameter). In non-square samples it is also illuminating to investigate the components $\langle D_x^2\rangle $ and $\langle D_y^2\rangle$ 
individually, to see how the lattice shape (and boundaries) affect the symmetry breaking. As will be demonstrated in the following sections, this issue is, 
in fact, of key importance for interpreting numerical results for non-square samples.

In the cylindrical semi-periodic systems it is useful to define the order parameter in such a way that the perturbing effects of the open edges are partially 
eliminated. As in Ref.~\onlinecite{jiang11}, for such systems with $L_x > L_y$ the summations in (\ref{dxsum}) and (\ref{dysum}) will normally be taken over 
only the central sites within a square of size $L_y \times L_y$.

In cases when the lattice coordinates $(x,y)$ are needed explicitly in the further discussion of correlation functions in the later sections, the numbering 
convention will be $x \in \{0,\ldots, L_x-1\}$ and $y \in \{0,\ldots, L_y-1\}$. 

\subsection{Convergence tests}

To examine the convergence properties of the projector method, the state (\ref{hmproj}) after $m$ operations with $H$ can be written in 
terms of eigenstates $|n\rangle$ of $H$ as
\begin{equation}
|\psi_m\rangle = \sum_{n} c_n E_n^m |n\rangle,
\end{equation}
where $c_n$ are the expansion coefficients of the trial state in the energy basis. Assuming that the ground state energy $E_0$ is the
largest in magnitude, $|E_0|\ge |E_n|,~\forall~n>0$, which is the case for sure with a Hamiltonian expressed using the singlet projectors 
(\ref{sproj}) and the signs as in (\ref{hj}) and (\ref{hq}), an expectation value of an operator $A$ not commuting with the Hamiltonian 
can be expanded as
\begin{equation}
\langle A\rangle_m = \langle 0|A|0\rangle + 2\langle 1|A|0\rangle \frac{c_1}{c_0}\left ( \frac{E_1}{E_0} \right )^m + \ldots .
\label{aconv1}
\end{equation}
Here $|1\rangle$ is the first excited state in the symmetry sector considered, which with an amplitude product state obeying all applicable
lattice symmetries is a singlet that is fully symmetric with respect to all the symmetry operations (translations, reflections, and rotations 
of the lattice). With the gap $\Delta=E_1-E_0$ and a large projection-power $m$, Eq.~(\ref{aconv1}) can be written as
\begin{equation}
\langle A\rangle_m = \langle 0|A|0\rangle + c \times {\rm exp} \left ( - \frac{m}{N} \frac{\Delta}{|e_0|} \right ),
\label{aconv2}
\end{equation}
where $e_0$ is the ground state energy per site, $e_0=E_0/N$, and $c$ is a constant. In order to achieve good convergence, one should therefore use a 
size-normalized projection power $m/N \gg 1/\Delta$. 

\begin{figure}
\includegraphics[width=8.25cm, clip]{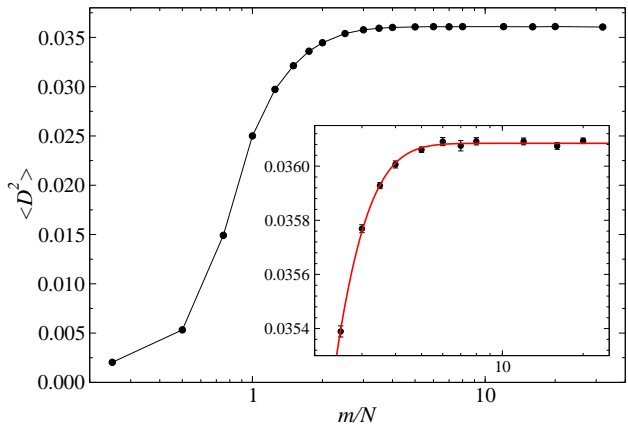}
\vskip-1mm
\caption{(Color online) The VBS order parameter as a function of the projection power $m$ (normalized by the system
size $N$) in simulations of the $Q_3$ model on a periodic $32\times 32$ lattice. The inset shows an exponential fit of
the form (\ref{aconv2}).}
\label{conv32}
\vskip-1mm
\end{figure}

The gapped VBS state being of primary interest here, $\Delta/\epsilon_0$ approaches a non-zero constant as the system size increases. One
may then expect good convergence properties with an essentially size independent $m/N$. However, for system sizes accessible in practice, the gap
still typically decreases significantly with the system size. In addition, the density of states above the gap increases as well. As
a consequence, $m/N$ has to be increased with the system size to ensure good convergence. Since the number of operations required for one full 
sweep of Monte Carlo updates of a configuration in the projector method is of order $m$,\cite{sandvik10b} the computation time in practice grows 
faster than $N$.

All results presented here were tested for convergence by carrying out several calculations with different projection powers 
$m/N \propto L$ (with $L={\rm max}[L_x,L_y]$ for non-square lattices) and making sure that there is no remaining detectable dependence on 
$m$. An example of a detailed convergence test is shown in Fig.~\ref{conv32}. Typically, $m/N=L/2$ was sufficient to ensure good convergence. 
In principle the singlet-singlet gap can be extracted by fitting the exponential form (\ref{aconv2}) to data such as those in Fig.~\ref{conv32} 
(as shown in the inset), but such gaps will not be studied here. 

\section{Extrapolation of VBS order}
\label{sec:extrap}

Previous ground-state and finite-temperature QMC studies have confirmed that both the $J$-$Q_2$ and $J$-$Q_3$ models, with the singlet projectors arranged as in 
Fig.~\ref{qterms}, have VBS-ordered ground state for large $Q_n/J$.\cite{sandvik07,melko08,jiang08,lou09} The maximal order parameter obtains for $J=0$ (pure 
$Q_n$ models) and, naturally, the order is more robust in the $Q_3$ model. The previous studies were mainly concerned with the critical and near-critical aspects 
of the N\'eel and VBS order parameters---the critical exponents as well as the emergent U($1$) symmetry seen in the VBS order parameter $(D_x,D_y)$. 

In this section some important aspects of the VBS order parameter will be discussed first, in particular the expected consequences of its emergent U($1$) symmetry. Then,
turning to numerical results, the magnitude of the VBS order parameter of the pure $Q_3$ model will be extracted first, to illustrate the convergence as a function 
of the lattice size for several cases of lattice shapes and boundary conditions. The $J$-$Q_2$ model, including the pure $Q_2$ case, is then considered in order to 
investigate potential problems arising when the VBS order is weaker. The quantum-critical scaling will also be discussed briefly, as it is directly related to the 
extrapolation problems when the VBS can be considered near-critical.

\subsection{Nature of the VBS order parameter}
\label{sec:vbsordernature}

Note first that the maximal columnar VBS order parameter is obtained for the state with no fluctuations in the valence bond basis---the state
with nearest-neighbor singlets on all bonds of every second column. If the singlets are oriented in the $x$ direction, then the order parameter components
defined in (\ref{dxsum}) and (\ref{dysum}) have the expectation values $\langle D_x\rangle=3/8$ (up to an arbitrary sign) and $\langle D_y\rangle=0$.
If the symmetry is not broken and the ground state is an equal superposition of the four degenerate valence-bond states with horizontal and
vertical bonds, then the expectation value of the squared VBS order parameter (\ref{d2def}) is $\langle D^2\rangle=\langle D_x\rangle^2 + \langle D_y\rangle^2=9/64$ 
in the limit of an infinitely large system. For finite systems there are corrections to this value, however, which are related to the non-orthogonality 
and over-completeness of valence bond states.\cite{beach06,tang11b} 

The emergent U($1$) symmetry property of the VBS order parameter \cite{senthil04a,senthil04b,sandvik07,kawashima07} and its related length-scale $\Lambda$ will be of importance in 
order to understand many of the results to be discussed here and in the later sections. For $L \ll \Lambda$, the order parameter $(D_x,D_y)$ on an $L \times L$ 
lattice behaves essentially as an isotropic 2D vector, while for $L \gg \Lambda$ the order parameter locks to one of the four angles $n\pi/2$. This is further 
discussed in Appendix \ref{appa}. Here, for $L_x \not= L_y$ lattices, with or without open edges, the U$(1)$--Z$_4$ cross-over will manifest itself also in how 
(on what length scale) the $90^\circ$ rotational symmetry of the VBS order parameter is broken on a lattice which does not have this symmetry. 

It should be noted that symmetry cross-overs such as the U$(1)$--Z$_4$ case discussed here also occur in many classical systems with dangerously irrelevant perturbations 
(i.e., ones that do not change the universality class of a phase transitions but reduce the degeneracy of the ordered state), e.g., the 3D XY-model with a $q$-fold 
symmetry-breaking field of the form $\cos(q\theta_i)$ (with $q\ge 4$).\cite{jose97,blankschtein84} There are several numerical studies of the scaling 
dimension of such a dangerously irrelevant perturbation and the nature of the cross-over and its length-scale 
$\Lambda$.\cite{caselle98,oshikawa00,carmona00,hove03,lou07} 

\begin{figure}
\includegraphics[width=7.5cm, clip]{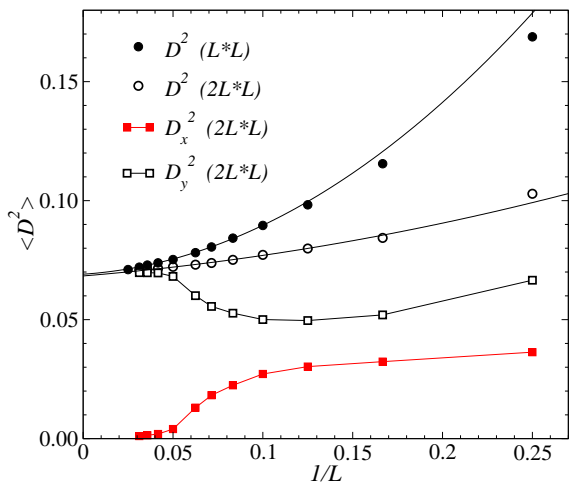}
\vskip-1mm
\caption{(Color online) Size dependence of the squared order parameter and its $x$ and $y$ components of the $Q_3$ model computed on periodic $L\times L$ 
and $2L \times L$ lattices. The curves passing through the $\langle D^2\rangle$ data are second-order polynomial fits (excluding the systems for which this form 
cannot be used). The error bars are much smaller than the plotting symbols (typically the standard deviation is $\approx 2\times 10^{-5}$).}
\label{perijq3}
\vskip-1mm
\end{figure}

\subsection{Strong VBS order in the pure $Q_3$ model}
\label{sec:jq3}

Fig.~\ref{perijq3} shows the size dependence of $\langle D^2\rangle$ of the $Q_3$ model computed on periodic $L\times L$ and $2L\times L$ 
lattices. For the latter systems the individual expectation values $\langle D_x\rangle^2$ and $\langle D_y\rangle^2$ are also shown (while these
are of course both equal to $\langle D^2\rangle/2$ for the $L\times L$ lattices). Here the convergence of $\langle D^2\rangle$ to a non-zero value
when $L \to \infty$ is apparent for both types of lattices. It is interesting to note that both the $x$ and $y$ components are nonzero on the $2L\times L$ 
lattices for small $L$, but for larger systems the symmetry is completely broken, eventually leading to $\langle D_x^2\rangle \to 0$, 
$\langle D_y^2\rangle \to \langle D^2\rangle$. Thus, on the non-square periodic lattices the columnar state with the bonds oriented parallel to the 
shorter lattice direction (here $L_y$) is energetically favored. This remains true also for larger aspect ratios $L_x/L_y$. 

The cross-over from partially broken to fully broken $x$-$y$ rotation symmetry, which in Fig.~\ref{perijq3} takes place for the $2L\times L$ systems for 
$L \approx 20$, should be related to the emergent $U(1)$ symmetry of the VBS order parameter.\cite{senthil04a,senthil04b,sandvik07} As discussed in Appendix 
\ref{appa}, for the $Q_3$ model no perfect U($1$) symmetry can be detected on periodic $L\times L$ lattices (since the length-scale $\Lambda$ is very short), but 
for a wide range of sizes the system is in a cross-over regime between U($1$) and Z$_4$ symmetry. The range of $L$ over which the cross-over to a purely 
$y$-ordered VBS takes place in Fig.~\ref{perijq3} is roughly where all traces of U($1$) symmetry vanish on the $L\times L$ lattices (as discussed in 
Appendix \ref{appa}).

\begin{figure}
\includegraphics[width=7.5cm, clip]{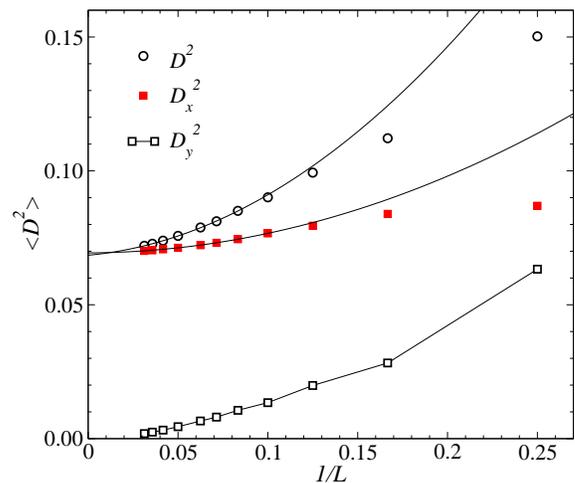}
\vskip-1mm
\caption{(Color online) Size dependence of the squared order parameters of the $Q_3$ model on cylindrical $2L \times L$ lattices (using the central 
$L\times L$ square for computing the expectation values). The smooth curves are second-order polynomials fitted to the $\langle D^2\rangle$ data 
for several of the largest system sizes.}
\label{openjq3}
\vskip-1mm
\end{figure}

Turning now to the quantitative behavior of the total order parameter for the largest systems in Fig.~\ref{perijq3}, as expected the order parameters for both 
lattice types extrapolate to the same value in the thermodynamic limit. Fits of the data for the largest systems to second-order polynomials are shown. Note, 
however, that this form is strictly not correct. For a discrete broken symmetry one would expect the asymptotic finite-size corrections to be exponentially 
decreasing with increasing system size. It is not easy to reach sufficiently large systems to observe this behavior, however. The second-order fits look reasonably 
good on the scale of the plot, but in fact they are not of high quality statistically when 6-8 data points are included. Including higher powers helps somewhat, 
but this can lead to fitted forms that do not behave monotonically as $1/L \to 0$. Such problems with the polynomial fits reflect a cross-over to the eventual 
exponentially rapid convergence. Using second-order fits for the largest few system sizes still should result in a reasonably accurate extrapolated order parameter. 
Normally such an extrapolation should give a lower bound on the actual value, but this cannot be guaranteed in the presence of statistical errors. In the case 
considered here, the results for $L\times L$ and $2L\times L$ extrapolate to $0.0691$ and $0.0684$, respectively, with the fits shown in Fig.~\ref{perijq3}.
Because of the issues with the, strictly speaking, wrong form of the fitting function, it is not meaningful to compute error bars on these numbers---the purely statistical
errors are smaller than the variations among fits with different polynomials and number of data points included. For the purposes of the investigations in this paper, 
the issue of statistical errors is only of minor importance, however (while the systematical errors due to a wrong fitting form are important).

\begin{figure}
\includegraphics[width=7.5cm, clip]{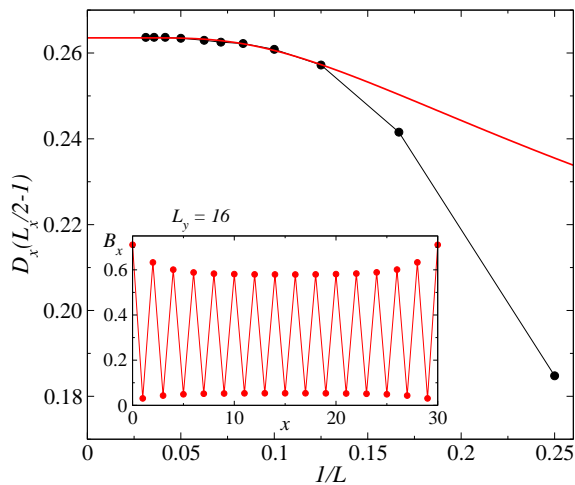}
\vskip-1mm
\caption{(Color online) Local columnar $x$ order parameter (\ref{dxxdef}) of the $Q_3$ model computed at the center of a $2L\times L$ cylinder.
The smooth curve is of the exponential form (\ref{aconv2}) and extrapolates to $0.264$. The inset shows the location dependent bond correlation 
function $\langle B_x(x)\rangle$ for a $32\times 16$ system.}
\label{dxjq3}
\vskip-1mm
\end{figure}

Data for cylindrical $2L\times L$ systems are shown Fig.~\ref{openjq3}. Here the order parameters are computed on the central $L\times L$ square.
In sharp contrast to the fully periodic $2L \times L$ systems, here it is the $x$ component of the order parameter that survives in the thermodynamic
limit. Thus, the open edges along the $y$ direction favor the bonds ordering perpendicularly to them, and this effect wins over the competing effect,
demonstrated in Fig.~\ref{perijq3}, of the aspect ratio $L_x/L_y>1$ favoring bonds ordering in the $y$ direction. Quadratic fits to $\langle D^2\rangle$ and 
$\langle D_x^2\rangle$ for a few of the largest system sizes extrapolate to $0.0685$ and 0.$0694$, respectively, in good agreement with the results for 
the periodic systems.

As a consequence of the open boundaries inducing an $x$-oriented VBS, the ordering pattern in this case is non-degenerate. Therefore, the
unsquared order parameter $\langle D_x\rangle$ is non-zero and should, in the thermodynamic limit, take a value agreeing with the squared order 
parameters extracted above; $\langle D_x\rangle \to \langle D^2 \rangle^{1/2}$. The expectation value of the nearest-neighbor spin correlator 
(\ref{bxdef}) indeed oscillates considerably as a function of the location along the $x$ direction, as shown in the inset of Fig.~\ref{dxjq3} 
for the $32 \times 16$ cylinder. The dimer order is clearly the strongest at the edges but remains large also in the interior of the system.

A local VBS order parameter for a system with bonds ordered along the $x$ axis can be defined as
\begin{equation}
D_x(x)=\langle B_x(x,y)\rangle - \half \langle B_x(x-1,y)+ B_x(x+1,y)\rangle,
\label{dxxdef}
\end{equation}
which is independent of $y$ on the semi-periodic cylindrical lattices (and can be averaged over $y$ in the QMC calculations). 
This quantity at the central column is shown as a function of the inverse system size in the main plot of Fig.~\ref{dxjq3}. Here an
asymptotic exponentially fast convergence can be seen clearly, which is illustrated with a fit to the form (\ref{aconv2}). This fit is of good
statistical quality and extrapolates to $0.264$, in good agreement with the values for $\langle D^2\rangle^{1/2}$ obtained above.
The magnitude of the  order parameter of the $Q_3$ model is, thus, $70\%$ of the largest possible value ($3/8$) for a columnar VBS.

\subsection{Reduced order in the $Q_2$ and  $J$-$Q_2$ models}
\label{sec:jq2}

\begin{figure}
\includegraphics[width=7.5cm, clip]{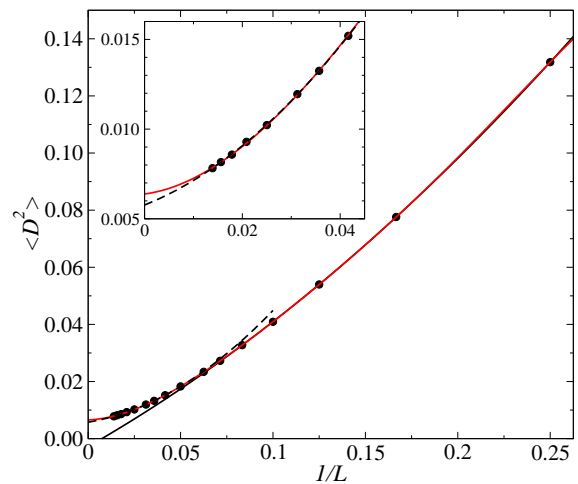}
\vskip-1mm
\caption{(Color online) Size dependence of the squared VBS order parameter of the pure $Q_2$ model on periodic $L\times L$ lattices.
The solid black curve in the main graph shows a fit of the $L\le 12$ data to a second-order polynomial (which extrapolates to an unphysical negative
value when $L\to \infty$). The solid red curve shows a $5$th-order polynomial fit to all the data, while the dashed black curve shows a 
quadratic fit to only the $L\ge 20$ data. The inset shows the behavior for the largest systems on a more detailed scale.}
\label{peri}
\vskip-1mm
\end{figure}

In the pure $Q_2$ model, the VBS order is considerably weaker than in the $Q_3$ model. The first study of this model gave the order parameter 
$D \approx 0.070$, or about $20\%$ of the maximal value, based on extrapolations of $L\times L$ results for $L\le 32$.\cite{sandvik07} While this 
order may still be regarded as quite strong, problems with extrapolating it correctly based on small to  moderate lattice sizes already start to 
become apparent. 

Fig.~\ref{peri} shows results for periodic $L\times L$ systems with $4 \le L \le 72$. A $5$th-order polynomial can be fitted very well to all these 
data and extrapolates to $0.0063$, about $10\%$ lower than the previous result. However, if only $L\ge 20$ data are used, a second-order polynomial is 
sufficient and the extrapolated value is significantly lower; $\langle D_x^2\rangle = 0.0058$. This illustrates the fact that polynomial fits based 
on small systems are not very reliable, because of the eventual exponential convergence (which is not yet fully apparent for the system sizes accessible). 
The resulting relative uncertainties are much larger than in the strongly ordered $Q_3$ model. The extrapolated value depends significantly on what 
system sizes are included in the fit and the order of the polynomial used. For the system sizes studied here, a pure exponential form does not yet work.

An important aspect of the finite-size scaling behavior in the $Q_2$ model is that the data for small to moderate lattices do not even clearly point to 
an ordered ground state. Fig.~\ref{peri} also shows a second-order fit to only the $L\le 12$ data points. The fit is statistically sound, but extrapolates
to a negative value. Without access to larger system sizes it is not possible, using fitting procedures like this in $1/L$, to determine whether the ground state of 
the infinite 2D lattice is ordered or disordered. At least $L=20$ is needed with $1/L$ extrapolations to definitely conclude that the ground state is ordered. 
It can be noted that an asymptotic $\propto 1/L^2$ behavior is expected if there is no long-range order, but this form should apply only for $L$ much 
larger than the correlation length. Note that the correlation length itself is also not easy to extract from the correlation functions unless $L \gg \xi$
(which is not the case here).

\begin{figure}
\includegraphics[width=7.5cm, clip]{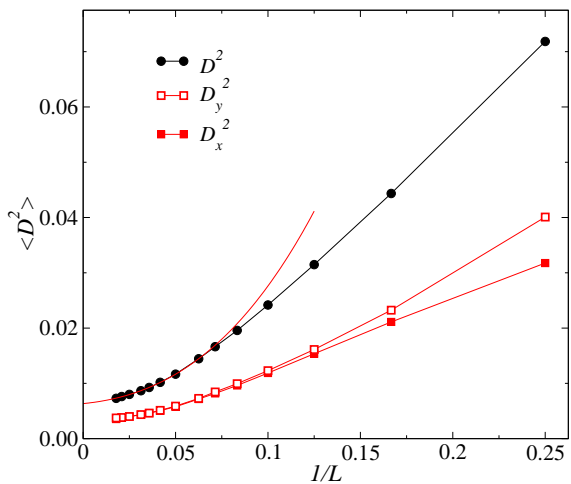}
\vskip-1mm
\caption{(Color online) Size dependence of the squared total VBS order parameter $\langle D^2\rangle$ and its individual $x$ and $y$ components of 
the pure $Q_2$ model on fully periodic $2L\times L$ lattices. The curve displayed for $1/L \le 0.125$ is a $4$th order polynomial fit to the $\langle D^2\rangle$ 
data for $L \ge 16$.}
\label{peri_2}
\vskip-1mm
\end{figure}

Fig.~\ref{peri_2} shows results for periodic $2L\times L$ lattices. Using polynomials to reliably extrapolate results to the infinite size limit is again
difficult. An example using a $4$th order polynomial with data for $L\ge 16$ is shown which extrapolates to $\langle D^2\rangle = 0.0063$. Here it is
again clear that the polynomial is not the correct form, because the fitted curve deviates significantly for the smaller systems not included in the fit.

The behavior of the individual $x$ and $y$ components in Fig.~\ref{peri_2} appears to be qualitatively different from that observed in the $Q_3$ model 
(Fig.~\ref{perijq3}). In the more strongly ordered $Q_3$ model the $y$ component is always significantly larger than the $x$ component, and for large 
systems it completely dominates (the $x$ component vanishing). In the $Q_2$ model, the length-scale $\Lambda$ of the cross-over from U($1$) to Z$_4$ symmetry 
is much larger, and the dimer order parameter acts as an essentially isotropic vector even for the largest lattices considered here. A cross-over to a behavior 
where the $x$ component vanishes (as in the $Q_3$ model) should take place for larger system sizes, but, according to the analysis for $L\times L$ lattices 
in Appendix \ref{appa}, the cross-over length is beyond what can currently be studied with QMC calculations, with there being only weak signals of a columnar 
state. Since the two components are almost equal in magnitude in Fig.~\ref{peri_2}, not knowing about the peculiar finite-size effects due to emergent 
U($1$) symmetry one may draw the erroneous conclusion from these data of the system being a plaquette VBS.

It is interesting to note in Fig.~\ref{peri_2} that the emergent $x$-$y$ symmetry is not manifested yet for the smallest systems. This reflects that fact that the 
continuous angular nature of the VBS order parameter only appears upon coarse-graining and $L < 10$ is not sufficiently large for representing a continuous VBS angle. 
The two cross-over length-scales, into and out of an U($1$) symmetric order parameter, have been investigated in detail in classical systems (clock models) exhibiting 
emergent U($1$) symmetry.\cite{hove03}

\begin{figure}
\includegraphics[width=7.5cm, clip]{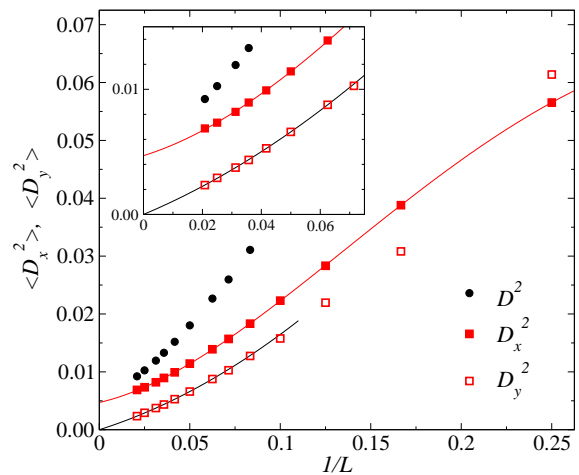}
\vskip-1mm
\caption{(Color online) Size dependence of the squared total order parameter $\langle D^2\rangle$ and its $x$ and $y$ components, computed for the 
$Q_2$ model on cylindrical $2L\times L$ lattices (including only the spins on the central $L\times L$ square in the definition of the order parameters). 
The curves are polynomial fits. In the case of $D_y^2$, no constant term was included. The inset shows the data for large systems on a more detailed scale.}
\label{open}
\vskip-1mm
\end{figure}

As in the $Q_3$ model, on the open-edge cylinders with $L_x>L_y$ the favored VBS ordering pattern is that with the bonds primarily in the $x$-direction. Fig.~\ref{open} 
shows results for $2L\times L$ cylinders. Here the effect of the edges to strongly favor $x$ ordering overcomes the tendency to U($1$) symmetry, and there is 
never any size range for which the $x$ and $y$ components are almost equal. Also here the behavior of both components for small lattices exhibit a naive extrapolation 
to a negative order parameter. For larger lattices $\langle D_x^2\rangle$ crosses over to a form extrapolating clearly to a non-zero value, while the $y$ component 
extrapolates to zero. A $4$th-order polynomial fit to all the $x$-component data gives $\langle D_x^2\rangle=0.0047$. This is significantly lower than the value quoted 
above for the examples of extrapolations of $L\times L$ data. However, the extrapolation is again sensitive to the lattice sizes included and the form of the fitting 
function used.

\begin{figure}
\includegraphics[width=7.5cm, clip]{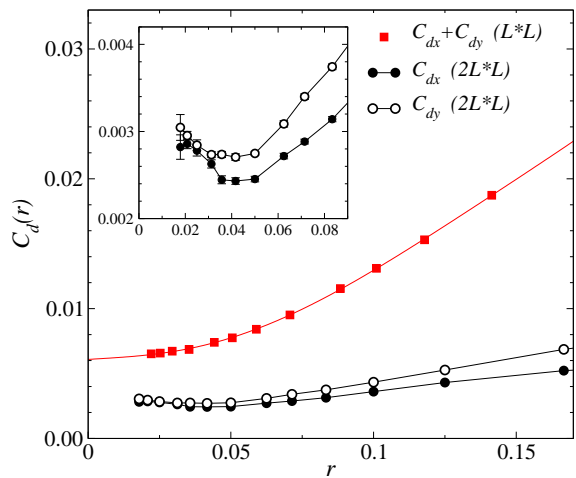}
\vskip-1mm
\caption{(Color online) Staggered component, Eq.~(\ref{dstar}) of the long-distance dimer correlations in the $Q_2$ model on periodic $L\times L$ 
and $2L\times L$ lattices. Here $r$ is the longest distance on the lattices; ${\bf r}=(L_x/2,L_y/2)$. The curve through the $L\times L$ data is a 
high-order polynomial fit. The inset shows the $2L\times L$ data on a more detailed scale.}
\label{longdist}
\vskip-1mm
\end{figure}

It is also useful to examine the long-distance VBS correlation function, which should contain less finite-size corrections to the infinite-size order
parameter than the sums over all correlations. The squared order parameter (\ref{d2def}) contains significant non-asymptotic contributions 
from short distances. Using the real-space dimer correlation function defined in Eq.~(\ref{cddef}), the staggered part in the case of the $x$ component 
(and an analogous form for the $y$ component) can be extracted according to
\begin{eqnarray}
&& C^*_{dx}(x,y)=\hbox{$\frac{1}{2}$}C_{dx}(x,y) \label{dstar} \\
&& \hskip17mm -\hbox{$\frac{1}{4}$}[C_{dx}(x-1,y)+C_{dx}(x+1,y)],\nonumber 
\end{eqnarray}
where a factor $1/2$ has been included in order for $C^*_{dx} (x\to \infty,y) \to \langle D_x^2\rangle$, with $D_x$ defined in Eq.~(\ref{cddef}), in 
the thermodynamic limit. Fig.~\ref{longdist} shows results for the longest distance on periodic $L\times L$ and $2L\times L$ lattices. For the $L\times L$ 
systems the sum of the $x$ and $y$ components is shown, along with a high-order polynomial fit that extrapolates to the infinite size order parameter 
$D^2=0.0061$. This extrapolation should be reasonably reliable, because the data for the largest systems flatten out clearly, reflecting the asymptotic 
exponential convergence (unlike the integrated quantity $\langle D^2\rangle$ in Fig.~\ref{peri}, where no flattening-out is yet seen). For the $2L\times L$ 
system no reliable extrapolation is possible, because both components exhibit non-monotonic behavior. The sum of the $x$ and $y$ correlations for large $L$ 
is nevertheless very close to the $L\times L$ results. 

It can also be noted in Fig.~\ref{longdist} that the individual components of the correlation function at long distance show somewhat less prominent 
$x$-$y$ symmetry than the integrated correlators in Fig.~\ref{peri_2}, although they are both still roughly equally large.  Again, in the thermodynamic limit 
one of the components, likely the $x$ component, will have to turn down and vanish, as the $L_x>L_y$ geometry favors ordering in the $y$ direction.

\begin{figure}
\includegraphics[width=7.5cm, clip]{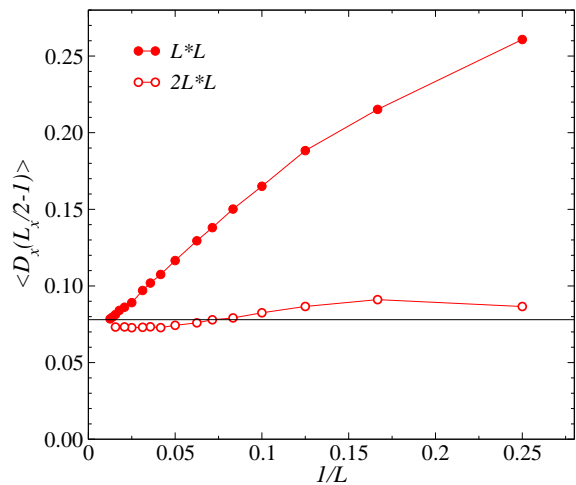}
\vskip-1mm
\caption{(Color online) Main graph: Open-edge induced order parameter of the $Q_2$ model at the center of cylindrical $L \times L$  and $2L \times L$ 
lattices. The horizontal line corresponds to the value of the infinite-size order parameter from the extrapolation in Fig.~\ref{longdist}.}
\label{dx00}
\vskip-1mm
\end{figure}

In the previous section, it appeared that the most reliable way to extract the order parameter in the thermodynamic limit is to exploit the symmetry-breaking 
open edges, using $\langle D_x(x)\rangle$ defined in Eq.~(\ref{dxxdef}). Fig.~\ref{dx00} shows such results for open-edge cylinders of size $L\times L$ as well
as $2L\times L$. Here the induced $x$ order appears to extrapolate to a value below the one obtained in Fig.~\ref{longdist} based on the long-distance
correlation function---for the $2L\times L$ systems $\langle D_x(L-1)\rangle$ is almost size independent for the largest systems, and one might hence conclude
that it has converged. The square of this value is $\langle D_x\rangle^2 \approx 0.073^2 \approx 0.053$, which seems too low compared to the results in 
Fig.~\ref{longdist}. 

The reason for this apparent inconsistency should again be related to the emergent U($1$) symmetry of the VBS order parameter: In addition to 
the $x$ component of the order parameter induced by the open edges, there still remains, for the accessible lattice sizes, a non-negligible $y$ component. 
This component is not locked-in by symmetry-breaking boundaries, however, but averages to zero if measured without first taking the square of its operator. 
The existence of a non-negligible fluctuating $y$ component nevertheless reduces the induced $\langle D_x\rangle$ from the full value, which should satisfy
$D^2 = \langle D_x \rangle^2 + \langle D^2_y \rangle$ for large systems. It is only when the system size exceeds the $U(1)$ length scale $\Lambda$ that one 
can expect the full order parameter to condense into the component $\langle D_x\rangle$, and this length scale cannot at present be reached for the $Q_2$ 
model. This shows again that the problem of extracting the VBS order parameter in the thermodynamic limit is a very delicate one. 

The examples shown here demonstrate that, when the VBS order is relatively weak (the length scale $\Lambda$ is large), it is 
important to look at the full order parameter, including both the $x$ and $y$ components. The long-distance correlation function (Fig.~\ref{longdist}) 
on $L\times L$ periodic lattices seems to be the fastest converging quantity, and it is in most cases best to use $L \times L$ lattices for extrapolations.
  
\begin{figure}
\includegraphics[width=7.5cm, clip]{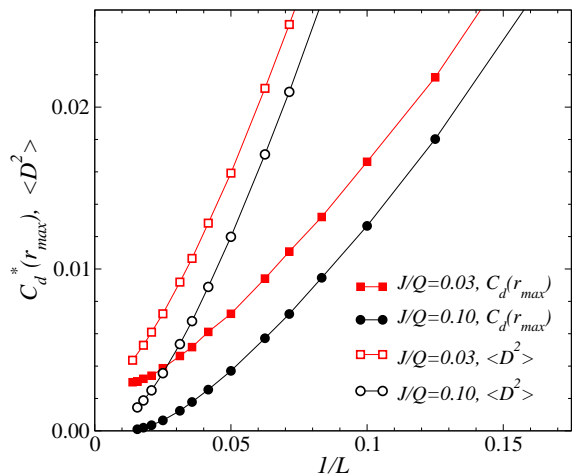}
\vskip-1mm
\caption{(Color online) The staggered part, Eq.~ (\ref{dstar}), of the long-distance correlation function (at $r_{\rm max}=\sqrt{2}L$) and the total dimer order parameter 
for the $J$-$Q_2$ model at $J/Q_2=0.03$ and $0.10$ on periodic $L \times L$ lattices.} 
\label{nearcrit}
\vskip-1mm
\end{figure}

When turning on the Heisenberg exchange $J$, the VBS order of the $J$-$Q_2$ model is reduced and vanishes when $J/Q_2 \approx 0.045$.\cite{sandvik10}
Here two cases are considered, $J/Q_2=0.03$ and $0.10$, with the latter corresponding to a near-critical N\'eel state. Fig.~\ref{nearcrit} shows results 
for the total squared VBS order parameter and the staggered part of the dimer correlation function (\ref{dstar}) averaged over the $x$ and $y$ directions.
With $\langle D^2\rangle$ graphed versus $1/L$ it is again difficult to extrapolate to infinite size based on small lattices. Here the lattices are nevertheless
sufficiently large for it to be apparent that the system at $J/Q_2=0.03$ is VBS ordered, while for $J/Q_2=0.1$ the decay is much more rapid and consistent with 
no VBS order. The corresponding long-distance correlations show these behaviors much more clearly, with the $J/Q_2=0.03$ data exhibiting the expected exponentially
fast convergence to a non-zero value for the largest sizes. Still, if data only for $L$ up to $\approx 10$ were available, it would not be possible to
unambiguously confirm the presence of long-range VBS order, even though the order parameter here is still above 10\% of the maximum value. 

Note that the long-distance correlation function decays exponentially as a function of $1/L$ in a non-VBS state, i.e., much faster than the $1/L^2$ behavior of
the total squared order parameter. It is therefore also much easier to confirm the absence of long-range order by studying the long-distance correlations.
 
\subsection{Quantum-critical scaling}
\label{sec:critical}

Ultimately, the difficulties in extrapolating the VBS order parameter to infinite size based on small systems will in many cases be related to 
critical scaling in the proximity of a quantum-critical point (or ``pseudo-critical'' scaling in cases where the transition out of the VBS state is
weakly first-order). A small system exhibits quantum criticality also slightly away from the critical point. Hence, data for a series of lattices 
may appear to extrapolate to a disordered state, even though the infinitely large 2D system is on the VBS side of a quantum phase transition. According 
to conventional finite-size scaling theory, the window around the critical point within which a system of linear size $L$ exhibits scaling is proportional 
to $L^{-1/\nu}$, where $\nu$ is the exponent governing the divergence of the correlation length. Depending on the prefactor, this window may be sizable 
for practically reachable lattice sizes. As will be shown next, this is one reason why fits to small-lattice data can give misleading results, e.g.,
in the case of $Q_2$-model results in Fig.~\ref{peri}. 

In addition to illustrating the near-critical VBS, the scaling of the N\'eel order parameter will also be briefly discussed here. According to past studies, 
both the $J$-$Q_2$ and $J$-$Q_3$ models are strong candidates \cite{sandvik07,lou09} for the deconfined quantum-criticality scenario,\cite{senthil04a} 
according to which both order parameters should be critical exactly at the same point. Results for the $J$-$Q_2$ model will be discussed here.

\begin{figure}
\includegraphics[width=8cm, clip]{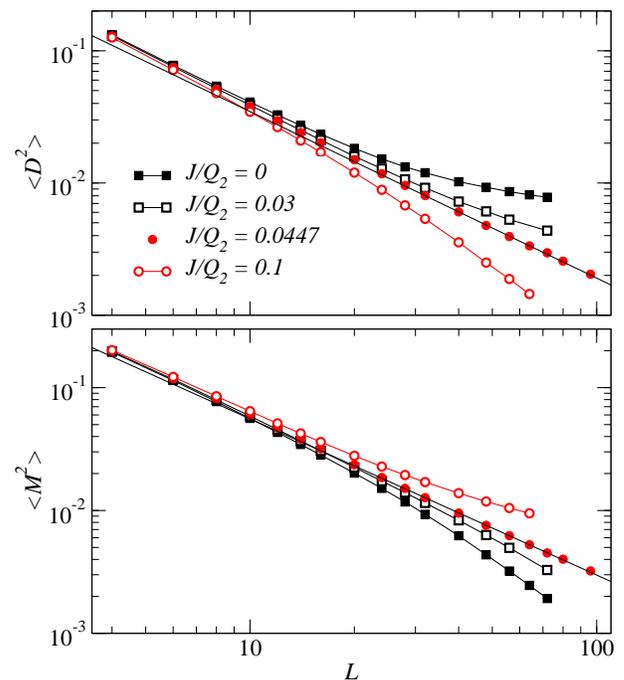}
\vskip-1mm
\caption{(Color online) Size dependence of the VBS (top) and N\'eel (bottom) order parameters of the $J$-$Q_2$ model at four different
coupling ratios. The point $J/Q_2=0.0447$ should be very close to the quantum-critical value according to the scaling analysis of the
spin stiffness carried out in Ref.~\onlinecite{sandvik10}. The straight lines fitted through the $J/Q_2=0.0447$ data (for system sizes $L\ge 32$)
have slope $-1.27$ in both cases.}
\label{critical}
\vskip-1mm
\end{figure}

While all numerical results so far are consistent with a single N\'eel--VBS transition point, it has proved remarkably difficult to determine the location 
$(J/Q_2)_c$ of this transition precisely. The most recent QMC studies point to a continuous transition with unusually large scaling corrections in the quantities 
normally used to extract the critical point, e.g., the spin stiffness and Binder cumulants.\cite{sandvik10,sandvik11,banerjee11b} These corrections 
have made it difficult to reliably extrapolate the critical coupling ratio $(J/Q_2)_c$ to infinite size. By using a logarithmic scaling correction to the
spin stiffness (which was not predicted in the original field-theory description of deconfined quantum-critical points but may appear with a modified action  
\cite{nogueira11}), $(J/Q_2)_c=0.0447 \pm 0.0002$ was obtained in Ref.~\onlinecite{sandvik10}. Using a conventional correction $\propto L^{-\omega}$, with 
small $\omega$ and a large prefactor (which potentially could be a consequence of the dangerously irrelevant operator responsible for the Z$_4$ symmetric VBS), 
gives a similar result. 

In Fig.~\ref{critical} the two order parameters are graphed versus the system size on log-log scales for coupling ratios close to the critical value. The N\'eel 
order parameter $\langle M^2\rangle$ (the squared sublattice magnetization) is the size-normalized $(\pi,\pi)$ Fourier transform of the spin correlation function 
(\ref{csdef}). Both order parameters indeed exhibit critical scaling at $(J/Q_2)=0.0447$. For other couplings the curves fan out in the way typical for critical points.

Interestingly, at $J/Q_2=0.0447$ both order parameters scale as $L^{-(1+\eta)}$ with $\eta \approx 0.27$ (with a purely statistical error bar of about $0.01$) when
$L\le 32$ systems are used in the fits. For the sublattice magnetization, this exponent is slightly smaller than in previous works,\cite{melko08,lou09} while the VBS 
exponent is somewhat larger than in Refs.~\onlinecite{sandvik07} and \onlinecite{lou09}. If these exponents are truly exactly the same, it would imply a duality of 
the effective low-energy field theory that had not been anticipated,\cite{senthil04a} but further detailed work, using larger system sizes and studying several 
coupling ratios in the neighborhood of $J/Q_2=0.0447$, will be required before such a claim can be made (and it could also be a coincidence that the two exponents 
are almost equal). Note also that the value of $\eta$ quoted here may also still be affected by sub-leading scaling corrections. 

For coupling ratios larger than the critical value, in Fig.~\ref{critical} exemplified by $J/Q_2=0.1$, the VBS order parameter turns downward, reflecting the faster 
decay to zero. Asymptotically, in the N\'eel state the decay should follow the $1/L^2$ form, but this can only be observed when the lattice size exceeds the correlation 
length (which is very large this close to the critical point). The sublattice magnetization turns upward, reflecting an extrapolation to a non-zero value. For smaller 
$J/Q_2$, here $0$ and $0.03$, the behavior is the opposite, reflecting a VBS state with no coexisting VBS order.

For the present purpose of detecting VBS order, an important aspect of the critical scaling is that, once a critical point has been identified, upward deviations 
from the power-law scaling, as seen in Fig.~\ref{critical} at $J/Q_2=0$ and $0.03$, still can demonstrate an ordered state when moving away from criticality.
It may be easier, in many models, to establish a critical point (or a first-order transition) than to accurately extrapolate the infinite-size value of the order 
parameter in a state with significant fluctuations (an order parameter significantly smaller than its maximum possible value). Based on the knowledge of the existence
of a phase transition, it may be possible to establish long-range order even in the presence of strong quantum fluctuations. This will be the case especially 
in calculations limited to much smaller systems. 

\section{Boundary symmetry breaking}
\label{sec:boundary}

One interesting aspect of the results presented in the previous section, exemplified in Figs.~\ref{perijq3} and \ref{openjq3}, is that the boundary conditions
dictate which of the order parameter components, $\langle D_x^2\rangle$ or $\langle D_y^2\rangle$, is the one surviving in the thermodynamic limit. For $90^\circ$ 
rotationally-symmetric periodic $L\times L$ lattices both order parameters are of course equal by symmetry (and spontaneous symmetry breaking in the thermodynamic 
limit will randomly select one of the directions), but in other cases only one of them should survive in the thermodynamic limit (i.e., the lattice shape acts 
like a symmetry-breaking field). Exactly how the symmetry is broken should be model dependent, and also dependent on fine details of the boundary conditions. 
Note that there are no ``neutral'' boundaries for a VBS, i.e., any boundary conditions should favor one component of the order parameter above the other
(expect perhaps for some unusual fine-tuned boundaries with adjustable couplings).

Here the $Q_2$ and $Q_3$ models will be used to illustrate the complexity of the boundary issues further, with direct measurements of the order-parameter 
components $\langle D_x\rangle$ and  $\langle D_y\rangle$ in systems where the edges break either the $x$-translational symmetry or both the $x$ and $y$ symmetries. 
The boundary effects are particularly interesting in view of the emergent U($1$) symmetry, due to which both order parameter components can survive up to 
large system sizes, as already shown in Sec.~\ref{sec:jq3} in the case of periodic systems. Here the ability of boundaries to twist the local order 
parameter in the $(D_x,D_y)$ plane will be studied.

\begin{figure}
\includegraphics[width=4.5cm, clip]{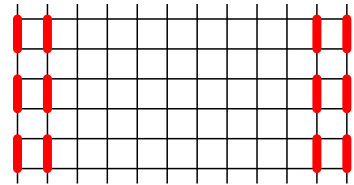}
\vskip-1mm
\caption{(Color online) A cylindrical lattice with modified open edges favoring columnar VBS order with vertical bonds.
The thick vertical bars represent $Q_2$ terms {\it excluded} from the summation in the Hamiltonian, Eq.~(\ref{hq}).}
\label{mlatt}
\vskip-1mm
\end{figure}

Two types of $2L \times L$ cylindrical lattices will be used. In addition to the case discussed so far, where the $y$-oriented edges are open and
uniform, a modified boundary that breaks the translational symmetry in the $y$ direction will also be studied. The modification acts as a field
inducing $D_y$ order at the edges. It is interesting to observe the interplay of this effect and the competing effect of the open boundary 
to lock in $D_x$ ordering when $L_x$ is even (as demonstrated in Fig.~\ref{dxjq3}). This aspect of the VBS ordering is also important in view of 
DMRG studies, where modified boundaries are often used.\cite{stoudenmire12,yan11,jiang11} Here the boundary modification will simply be accomplished 
by {\it excluding} from the Hamiltonian the $Q_2$ or $Q_3$  terms with vertical bonds closest to an edge on every second row, as illustrated in 
Fig.~\ref{mlatt} in the case of $Q_2$ terms. Results obtained with only one of the edges modified will be compared with the case of both edges 
modified in the same way.

The local variations of the VBS vector order parameter $(D_x,D_y)$ of the $J$-$Q_2$ model were previously investigated for $L\times L$ lattices 
with all open edges.\cite{kaul08} The formation of a vortex-like structure in the order parameter was noted. In the cases studied here, there is
still translational symmetry with period two along the $y$-axis and, therefore, a 1D description of the order parameter as a function of
the $x$ coordinate suffices. The local $x$ and $y$ order parameters are defined using the dimer operator $B_{\hat x}$ in Eq.~(\ref{bxdef});
\begin{eqnarray}
&&D_x(x)=\bigl [\langle B_x(x,y)\rangle-\half \langle B_x(x-1,y)\rangle \label{dxvsxdef}  \\ 
   && \hskip18mm -\half \langle B_x(x+1,y)\rangle \bigr ](-1)^x, \nonumber \\
&&D_y(x)=\bigl [\langle B_y(x,y)\rangle-\langle B_x(x,y+1)\rangle \bigr ](-1)^y \label{dyvsxdef} .
\end{eqnarray}
These quantities are independent of the $y$ coordinate (and an average is taken in the simulations to improve the statistics). A VBS angle $\theta(x)$
can also be defined,
\begin{equation}
\theta(x)={\rm atan}\left ( \frac{D_y(x)+D_y(x+1)}{2D_x(x)} \right ),
\label{thetadef}
\end{equation}
such that $\theta=0$  and $\theta=\pi$ for a fully $x$ or $y$ oriented VBS order, respectively. The reason for using the sum
$D_y(x)+D_y(x+1)$ in the numerator under atan$()$ is that an $x$-oriented column of bonds labeled by $x$ is located between the
$y$-oriented columns at $x$ and $x+1$ (although such a detail of the definition of the local angle is not strictly important, 
and there are other equally good definitions giving the same result for large systems).

\begin{figure}
\includegraphics[width=7.5cm, clip]{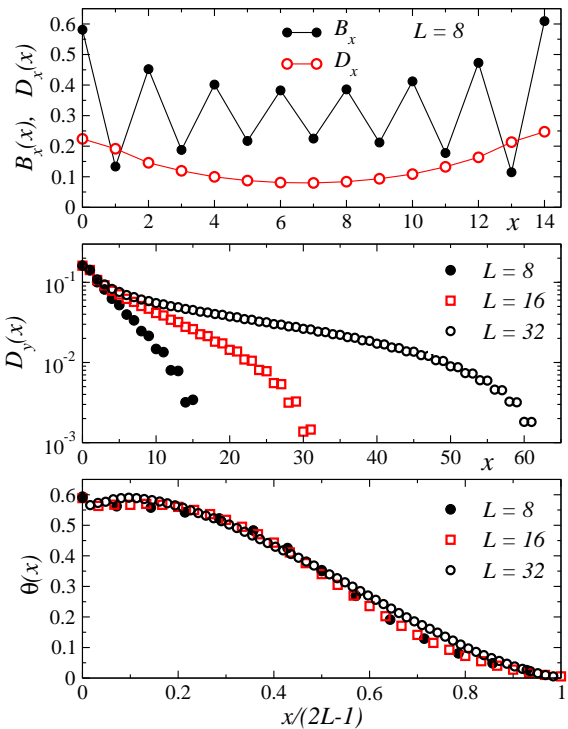}
\vskip-1mm
\caption{(Color online) Location dependent expectation value of the VBS order parameter of the $Q_2$ model on $2L\times L$ cylinders with the
left $(x=0)$ edge modified by the symmetry-breaking perturbation (inducing $y$-oriented order) illustrated in Fig.~\ref{mlatt}. The right edge is 
kept uniform. The top panel shows both the bare dimer expectation value $\langle B(x)\rangle$ and the dimer order parameter $\langle D_x(x)\rangle$ 
extracted from it according to Eq.~(\ref{dxvsxdef}) for $L=8$. The middle panel shows the $y$ order parameter defined according to Eq.~(\ref{dyvsxdef}) 
for $L=8,16$, and $32$. The bottom graph shows the VBS angle extracted from the $x$ and $y$ order parameters according to Eq.~(\ref{thetadef}).} 
\label{edg1q2}
\vskip-1mm
\end{figure}

Fig.~\ref{edg1q2} shows results for the $Q_2$ model with only one modified edge. Oscillations in the bare dimer expectation value $\langle B_{x}(x)\rangle$ are 
present (top panel as in the case of the uniform edge in Fig.~\ref{dxjq3}. In this case, however, the function is not reflection symmetric, due to the unequal left and right 
edges of the cylinder. The order parameter $D_x(x)$ is the largest at the edges. Away from the edge it decays toward a value at the center of the system which 
is somewhat smaller than the locked-in order parameter previously extracted based on the data in Fig.~\ref{open} (which can be seen by analyzing data for 
several system sizes, not shown here). This is because the modified edge also leads to some amount of $y$ order (middle panel of Fig.~\ref{edg1q2}, and although 
this induced order decays rapidly when moving away from the modified edge it does not go away completely, even close to the opposite edge.

The rather smooth decay of the $y$ order to almost zero at the opposite edge can be explained as due to the open edge strongly favoring $x$ ordering in its 
vicinity, even with the modification that breaks the $y$ translational symmetry. The modified edge therefore induces both $x$ and $y$ order, i.e., the VBS angle 
(\ref{thetadef}) is $0 < \theta < \pi/2$. Since the second edge does not break the $y$ translational symmetry explicitly, the $x$ ordering can completely dominate there, 
leading to a very small $\langle D_y\rangle$. The smooth transition from mixed $x$ and $y$ to almost pure $x$ order is seen clearly in the VBS angle graphed 
in the bottom panel of Fig.~\ref{edg1q2}. Away from the edges, the total order parameter for large systems, $D=(\langle D_x\rangle^2+\langle D_y\rangle^2)^{1/2}$, 
approaches the value extracted for this model in the previous section. A maximum in the angle develops with increasing size close to the modified edge.

\begin{figure}
\includegraphics[width=7.6cm, clip]{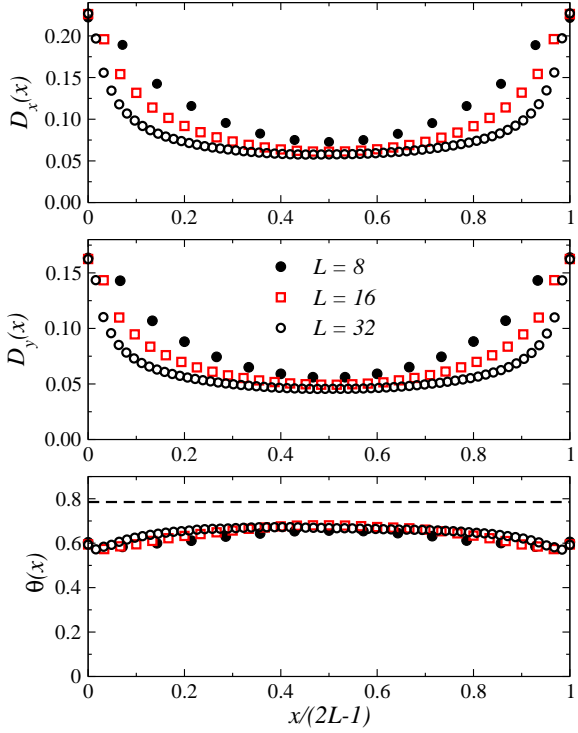}
\vskip-1mm
\caption{(Color online) Location dependent expectation values of the $x$ and $y$ VBS order-parameter components, Eqs.~(\ref{dxvsxdef}) and 
(\ref{dyvsxdef}), of the $Q_2$ model on $2L\times L$ cylinders with both edges modified by a symmetry-breaking perturbation (favoring $y$-oriented 
bond order). The corresponding VBS angle extracted using Eq.~(\ref{thetadef}) is shown in the bottom panel. The horizontal dashed line is at $\Theta=\pi/4$ 
(corresponding to equal $x$ and $y$ order parameter parameters).} 
\label{edg2q2}
\vskip-1mm
\end{figure}

Fig.~\ref{edg2q2} shows results for systems with the $y$ symmetry broken at both edges. Also in this case it would appear that both the $x$ and $y$ order parameters 
survive throughout the whole system in the thermodynamic limit. Convergence of both components as well as the angle at the center of the system is seen. The VBS angle 
here being only slightly less than $\pi/2$ corresponds to an almost equal mixture of $x$ and $y$ order.

\begin{figure}
\includegraphics[width=7.5cm, clip]{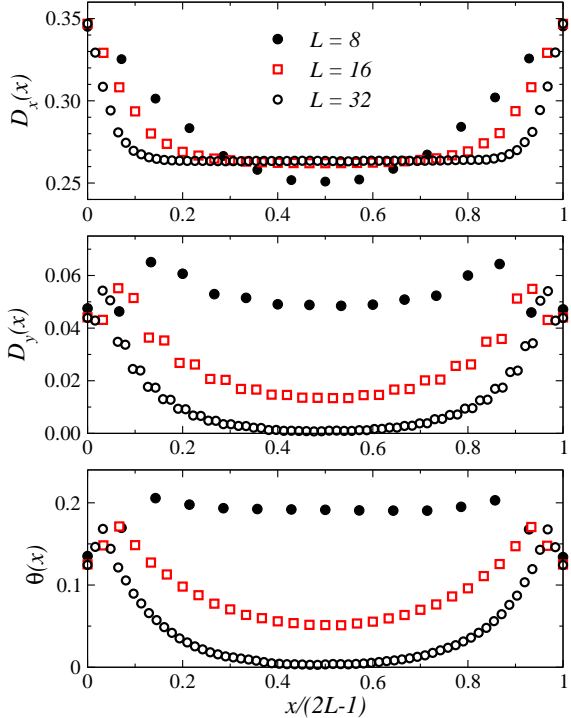}
\vskip-1mm
\caption{(Color online) Same as Fig.~\ref{edg2q2} for the $Q_3$ model.}
\label{edg2q3}
\vskip-1mm
\end{figure}

In spite of the apparent convergence of the VBS angle to a value close to $\pi/2$ in Fig.~\ref{edg2q2}, the survival of both $x$ and $y$ order in the thermodynamic 
limit due to the modified edge is illusory. Since the VBS order is columnar, eventually, for very large systems, one would expect only $x$ or only $y$ order to survive.
The explanation of the behavior seen is again the very large U($1$)--Z$_4$ cross-over scale in the $Q_2$ model (as discussed in Appendix \ref{appa}). It is then
interesting to look at the same quantities in the $Q_3$ model, where there is no clear U($1$) symmetry (as also shown in Appendix \ref{appa}), i.e., the length-scale $\Lambda$ 
is very short in this case. Results analogous to those in Fig.~\ref{edg2q2} for the $Q_2$ model are shown in Fig.~\ref{edg2q3} for the $Q_3$ model. In this case, one can 
see clearly how the $y$ component vanishes with increasing system size away from the edges, while the $x$ order stabilizes to a constant value. Since the $x$ component is 
the surviving one, its approach to its bulk value should be governed by the standard VBS correlation length $\xi$. The decay of the $y$-component should reflect $\Lambda$, 
however (since the presence of $y$ order is due to the angular twisting of the order parameter). This is a direct physical method to access the U($1$)  length-scale, 
providing an attractive alternative to studying the order-parameter distributions discussed in Appendix \ref{appa}.

\begin{figure}
\includegraphics[width=7.5cm, clip]{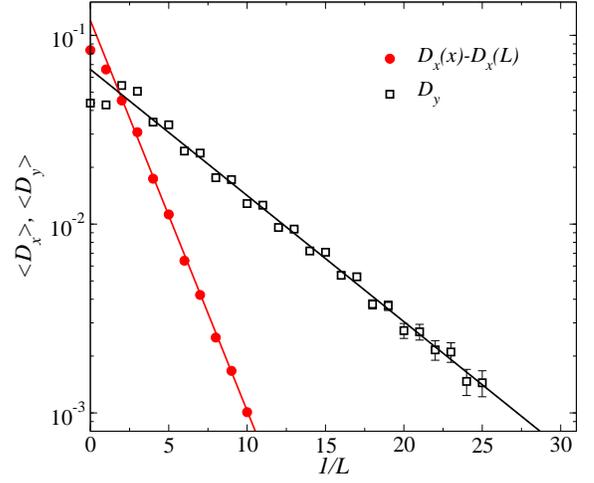}
\vskip-1mm
\caption{(Color online) The $x$ and $y$ components of the induced order parameter close to a modified edge of the $Q_3$ model on a $128\times 64$ lattice.
These data are the same as those shown in the top ($x$) and middle ($y$) panels of Fig.~\ref{edg2q3} for smaller systems, but with the non-zero constant behavior 
at the center of the system subtracted off in the case of the $x$ component. The lines are exponential fits, giving decay lengths $1.9$ and $6.5$ for the 
$x$ and $y$ components, respectively.}
\label{xydecay}
\vskip-1mm
\end{figure}

The decays of the two componentss are analyzed quantitatively for a larger system in Fig.~\ref{xydecay}. Excluding the points immediately adjacent to the edge, the decays 
are of almost pure exponential form (with an even-odd effect seen for the $y$ component), giving $\xi=1.9$ extracted from the $x$ component and $\Lambda=6.5$ from 
the $y$ component. A similar analysis for the $Q_2$ model (not shown here), based on systems with up to $128\times 64$ sites, gives $\xi \approx 25$ (and $\Lambda$ much
larger still), but this estimate is not reliable because the form of the decay is affected by the proximity to the critical point and is far from a pure exponential at 
the accessible distances. Larger system sizes are required in this case, especially for extracting $\Lambda$, which is larger than 100 lattice constants according
tho the analysis in Appendix \ref{appa} (perhaps being several hundred lattice constants). A systematic study of the divergence of the decay lengths of the $J$-$Q_3$ 
model upon approaching the quantum-critical point will be presented elsewhere.

\section{Long cylinders}
\label{sec:cylinder}

In the previous sections the 2D limit was approached in systems with fixed aspect ratio $L_x/L_y$. In principle the limit can also be accomplished with one of the 
lengths taken to infinity first, e.g., $L_x \to \infty$ for fixed $L_y$ and then $L_y \to \infty$. The behavior of the long-distance correlation functions, and, 
therefore, the squared VBS order parameter $\langle D^2\rangle$, should not necessarily be expected to be smooth, however. Although VBS ordering amounts to breaking 
a discrete symmetry, and order can therefore, in principle, exist for any $L_y$ in the infinitely long 1D cylinder geometry, the survival of the order for small 
$L_y$ is not guaranteed. Clearly, there will be enhanced fluctuations associated with the 1D nature of these systems, which may destroy the ground-state order of a
Hamiltonian exhibiting long-range VBS order in the 2D limit.

\begin{figure}
\includegraphics[width=7.5cm, clip]{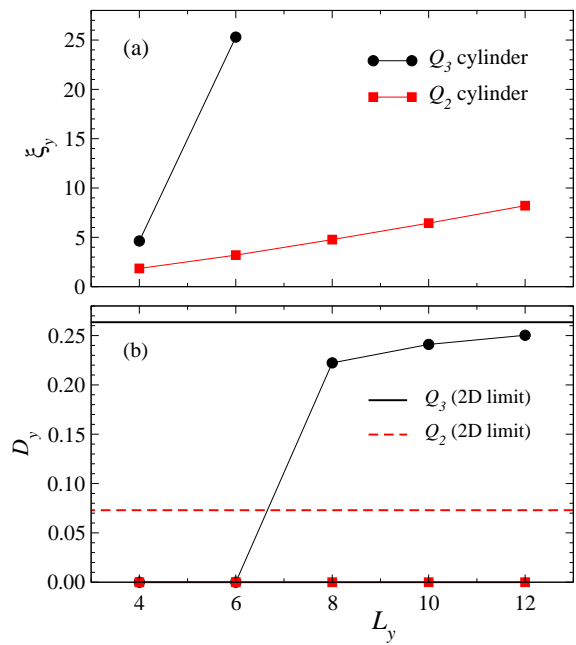}
\vskip-1mm
\caption{(Color online) VBS order parameter and correlations lengths on infinite ($L_x\to \infty$)  $Q_2$ and $Q_3$ cylinders of width
$L_y=4,6,8,10,12$. (a) The correlation length extracted using the dimer correlations with $y$-oriented bonds as a function of $L_y$ for those 
cylinders that have disordered ground states (the $x$ correlation lengths are about $5-10\%$ smaller). (b) The order parameter versus $L_y$, 
along with the corresponding 2D order parameters (shown with the horizontal lines).}
\label{cylsummary}
\vskip-1mm
\end{figure}

A well known system with discrete symmetry breaking is useful for illustrating the potentially unsmooth 1D to 2D cross-over: The Ising model 
with nearest-neighbor coupling $J_z$ in a transverse magnetic field $h_x$ has a phase transition to an ordered (in the $z$ spin direction) 
state at a critical value $(h_x/J_z)_c$. On a 1D linear chain the critical ratio is $(h_x/J_z)_c=1$, while on the 2D square lattice 
it is $(h_x/J_z)_c\approx 3.05$.\cite{hamer00} For an $L_x\times L_y$ lattice with $L_x\to \infty$ one can expect $(h_x/J_z)_c$ 
to be a monotonic increasing function of $L_y$. Therefore, for a fixed field $1 < h_x/J_z < 3.05$, one can expect cylinders with small $L_y$ to 
be disordered, while above some ``critical'' $L_y$ the system will be ordered. One can expect the same kind of behavior of a 2D VBS as well, when 
restricting it to a finite cylinder, unless the 2D order parameter is extremely large so that even the smallest cylinder remains in the 
ordered phase.

In the discussion below, only cylinders of even $L_y$ will be considered, so that the lattice is commensurate with columnar VBS order
in both the $x$ and $y$ direction. $J$-$Q$ models with odd $L_y$ cannot be studied with the QMC method used here, because of sign
problems arising due to geometric frustration of the spin interactions.

\subsection{Destruction of VBS order on cylinders}
\label{sec:cyldestruct}

As shown in Sec.~\ref{sec:jq3}, the ground state of the pure $Q_3$ model is strongly VBS ordered, the order parameter being at $70\%$ 
of the maximum possible value. One might expect this to be sufficient for the order to be stable also on thin cylinders when $L_x\to \infty$. 
However, it turns out that such cylinders of width $L_y=4$ and $6$ are disordered, while for $L_y=8$ and above the order parameter is already close to 
the 2D limiting value. For the pure $Q_2$ model, where the 2D order parameter is about $20\%$ of the maximum value, no order was found on $L_x \to \infty$ 
cylinders with $L_y$ up to $12$. Larger widths were not studied due to prohibitively long computation times. The results for both models are summarized 
in Fig.~\ref{cylsummary}. The results underlying these conclusions are discussed next.

\begin{figure}
\includegraphics[width=8cm, clip]{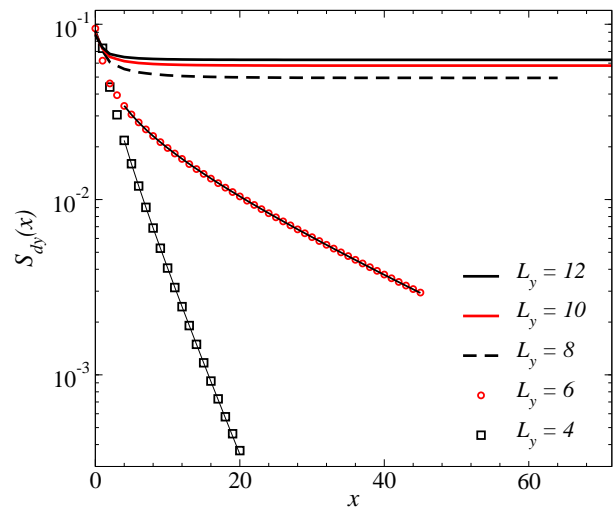}
\vskip-1mm
\caption{(Color online) VBS correlation functions, as defined in Eq.~(\ref{sydef}), for $y$-oriented dimers in the $Q_3$ model as a function of the separation 
in the $x$-direction on cylinders in the $L_x \to \infty$ limit. For $L_y=4,6$, fitted curves of the form $C \propto {\rm exp}(-x/\xi)/x^\alpha$ 
to the $x \ge 4$ data are also shown (with $\alpha \approx 0.5$ in all cases).}
\label{jq3cx}
\vskip-1mm
\end{figure}

It is useful to define correlation functions that are averaged over the short ($y$) direction. The following functions,
based on the definition (\ref{cddef}) of the elementary dimer correlator, can be used to detect columnar VBS order with the bonds 
oriented either along the $x$ or the $y$ direction;
\begin{eqnarray}
&& S_{dx}(x)=\frac{1}{L_y}\sum_{y=0}^{L_y-1}\bigl [C_{dx}(x,y) \label{sxdef} \\ 
     && \hskip15mm -\half C_{dx}(x-1,y)-\half C_{dx}(x+1,y)\bigr ], \nonumber \\
&& S_{dy}(x)=\frac{1}{L_y}\sum_{y=0}^{L_y-1}C_{dy}(x,y)(-1)^y. \label{sydef}
\end{eqnarray}
Here it is appropriate to use periodic boundary conditions in both lattice directions. In order to achieve the limit $L_x \to \infty$, aspect
ratios $L_x/L_y$ up to $32$ were studied for $L_y$ up to $12$.

In the $Q_3$ model, the $y$-dimer correlator $S_{dy}(x)$ approaches a non-zero constant for large $x$ when $L_y \ge 8$, as shown in 
Fig.~\ref{jq3cx}, while for $L_y=4,6$ the correlations decays exponentially with distance. The behavior is not purely exponential but follow
the form $S_{dy}(x) \propto x^{-\alpha}{\rm exp}(-x/\xi)$, with $\alpha \approx 0.5$. This form with $\alpha=1/2$ is the
Ornstein-Zernike (mean-field) form expected in a $d=(1+1)$ dimensional system, where $\alpha=(d-1)/2$. The correlation lengths extracted from fits 
to this form (with $\alpha$ regarded as a free parameter, to produce somewhat better fits) are shown in Fig.~\ref{cylsummary}(a). The $x$-oriented 
correlation function $S_{dx}(x)$ is exponentially decaying for all $L_y$, i.e., these systems are purely $y$-ordered in the thermodynamic limit 
(as was also found in Sec.~\ref{sec:jq3} for periodic $2L\times L$ systems when $L \to \infty$). For $L_y=4,6$ the $x$ correlation lengths are 
slightly smaller than the $y$ ones. The $y$ correlation lengths are graphed in Fig.~\ref{cylsummary}(a).

\begin{figure}
\includegraphics[width=8.4cm, clip]{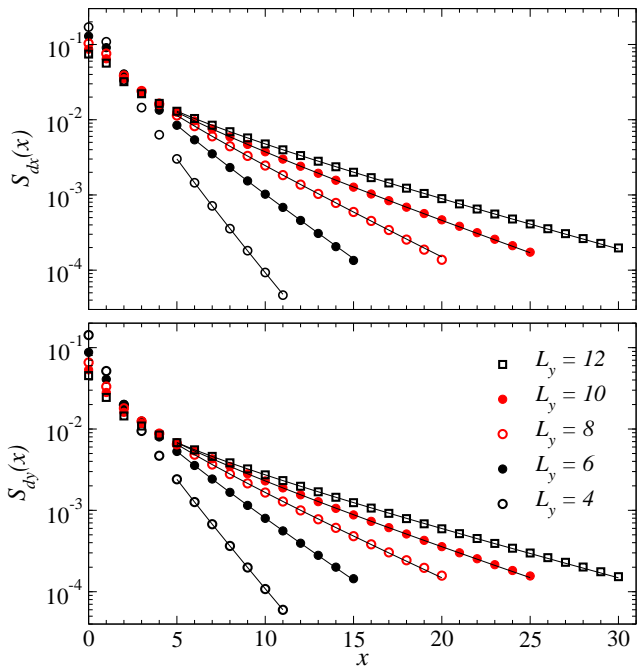}
\vskip-1mm
\caption{(Color online) VBS correlation functions, as defined in Eqs.~(\ref{sxdef}) and (\ref{sydef}), for $x$- (top panel) and $y$-oriented (bottom panel) dimers
in the $Q_2$ model as a function of the separation in the $x$-direction on cylinders in the $L_x \to \infty$ limit. Fits of the data for $x>4$ to the form 
$S \propto {\rm exp}(-x/\xi)/x^\alpha$ (woth $\alpha\approx 0.5$ in all cases) are shown as solid curves.}
\label{jq2cx}
\vskip-1mm
\end{figure}

\begin{figure}
\includegraphics[width=7.75cm, clip]{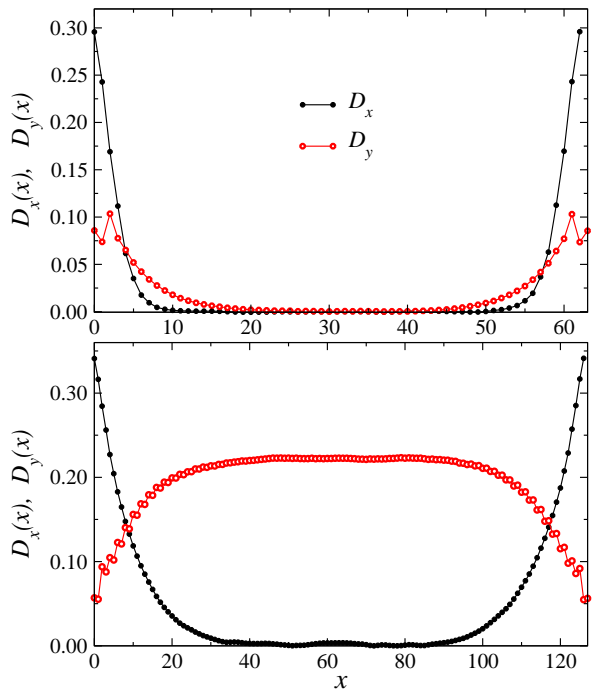}
\vskip-1mm
\caption{(Color online) Boundary induced $x$ and $y$ components of the dimer parameter of the $Q_3$ model on $64\times 4$ (top)
and $128 \times 8$ (bottom) lattices. Both edges are modified to induce $y$ order, as discussed in Sec.~\ref{sec:boundary}.}
\label{longdimx}
\vskip-1mm
\end{figure}

In Fig.~\ref{longdimx} both the $x$ and $y$ correlation functions for the $Q_2$ model are graphed for all even-width cylinders with $L_y=4,\ldots,12$, along 
with fits to the exponential form discussed above. The $y$ correlation length $\xi_y$ is the larger one (about $5-10\%$ larger than $\xi_x$) and is graphed 
versus $L_y$ in Fig.~\ref{cylsummary}(a). The correlation length grows roughly linearly with $L_y$ for these cylinders. It would be interesting to go to 
even larger $L_y$ to study the form in greater detail, and of course to find the threshold width for ordering in this case (where presumably the correlation 
length should diverge, if one regards $L_y$ as a continuous parameter). The rather small correlation lengths for $L_y$ up to $12$ suggest that it may be 
difficult to reach the critical width with QMC calculations at present.

The destruction of the VBS order even on rather wide cylinders is surprising. In the $Q_3$ model, judging by the decay of the $x$ component of the order parameter
in Fig.~\ref{xydecay}, the 2D correlation length is approximately $2$ lattice constants. The lower width $L_y=8$ for ordering on infinitely long cylinders is therefore 
roughly four times the correlation length. Moreover, related to the short correlation length, the 2D order parameter is as large as $70\%$ of the classical value. One 
might have expected such a system to be describable essentially in terms of classical (orthogonal, hard-core) dimers with quantum fluctuations of the nature present 
in quantum dimer models. It has been expected that a VBS under these conditions should be ordered even on narrow cylinders.\cite{yao11} The results obtained here suggest 
that the non-orthogonality of the singlets (the true quantum dimers) has a dramatic effects of reducing the order on cylinders, in contrast to this effect actually enhancing 
the dimmer-dimer correlations relative to those in corresponding dimer models in critical 2D systems.\cite{tang11b,albuquerque10} On the other hand, to the author's knowledge, 
quantum dimer models that order in the 2D limit \cite{alet05} have actually not been extensively studied in long-cylinder geometry. Such studies would clearly be worthwhile, 
in light of the surprising results obtained here.

In the $Q_2$ model the correlation length should be in the range $20 \sim 30$ (with, as already discussed above, the large uncertainty being due to the fact 
that system sizes $L \gg \xi$ are needed to determine $\xi$ accurately), and one can, thus, expect, roughly, $L_y \approx 100$ to be needed before ordering 
sets in on the cylinders in this case.

One might speculate that the emergent U($1$) symmetry could play some role in destroying the VBS order on the long cylinders. The local coarse-grained VBS order parameter 
$(D_x,D_y)$ is an essentially isotropic 2D vector up to a large length scale $\Lambda \sim \xi^{1+a}$ with $a>0$ (with the best estimate so far \cite{lou09} being 
$a=0.20 \pm 0.05$). If the order parameter were truly a vector with isotropic angular fluctuations, long-range order on the 1D $L_x \to \infty$ cylinders would 
be strictly prohibited.\cite{mermin66} The almost continuous order parameter could then be argued to contribute to the loss of order. If so, one would expect a 
critical state to replace long-range order, however, of which there are no signs here---the VBS order decaying exponentially starting from short distances. 
There is no cross-over from a critical behavior, which might have been expected if almost U($1$) symmetric angular VBS fluctuations were responsible for the destruction 
of long-range order. The role of emergent U($1$) symmetry on cylinders is nevertheless interesting and should be studied more systematically in the future.

Regardless of the exact relationship between the 2D correlation length and the ordering threshold on cylinders, the very short correlation lengths found in the 
$Q_2$ model (ranging from about $\xi \approx 2$ for $L_y=4$ to $\xi \approx 8$ for $L_y=12$) show the dangers of using the long-cylinder geometry for drawing 
conclusions about the presence or absence of VBS order in the 2D limit. Order likely appears in the $Q_2$ model, and probably in most models for which the existence 
of VBS order is under debate, for $L_y$ far exceeding the maximum size that can currently be studied (especially if QMC methods cannot be used and DMRG would be 
the best choice of method). 

The above conclusions regarding ordered and disordered cylinders reached based on correlation functions in long periodic $L_x \times L_y$ systems can also be 
confirmed by examining open-edge cylinders, in which a unique VBS can be locked in for even $L_x$ (as discussed in the case of $2L\times L$ cylinders in 
the preceding sections). Fig.~\ref{longdimx} shows results for longer $Q_3$ cylinders in which the boundary perturbation inducing $y$ order was also 
applied at both edges (as in Fig.~\ref{mlatt}). For $L_y=4$, both order parameter components decay quickly away from the edges, while for $L_y=8$ the $y$ 
component stabilizes at the center of the system, at a value agreeing with that extracted on the basis of the correlation functions [shown in 
Fig.~\ref{cylsummary}(a)]. Here, although the open edges favor $x$ order more than the perturbations favor $y$ order, the $y$ component eventually
wins because that is the component favored just by having a finite $L_y$, and this effect scales with $L_x$. In contrast, for the $2L\times L$ cylinders 
with the same types of edges, it is the $x$ component that survives in the thermodynamic limit, as seen in Fig.~\ref{edg2q3}.

\section{Conclusions and discussion}
\label{sec:summary}

\subsection{General summary and conclusions}

Several bench-mark results for the finite-size behavior of the VBS order parameter have been presented in this paper. The $J$-$Q$ and pure $Q$ models 
allowed investigations of both strongly and weakly ordered ground states. The main general conclusion (which should be valid for VBS states in many
systems) drawn from these studies is that even when the VBS order is relatively strong on the infinite 2D lattice (e.g., $10-20\%$ of the maximum value 
attainable), results for small and moderate lattices (e.g., with up to hundreds of spins) can exhibit nearly critical behavior. The squared VBS order 
parameter then appears to extrapolate to zero in the thermodynamic limit. In the $J$-$Q$ model, this behavior can be traced to a rather large 
quantum-critical scaling regime around the critical value of $J/Q$, where the behavior follows closely that obtaining at a critical point. 

The extrapolation to infinite size may at first sight seem easier when symmetry-breaking boundaries are used (as is often done in the context of 
DMRG studies\cite{white07}), so that the order parameter can be computed directly (having a considerably larger value than its square when the VBS order 
is not very strong). However, a small order parameter ($10-20\%$ of the maximum value in the VBS systems considered here) is very difficult to 
extrapolate accurately in this way, partially because the symmetry is not completely broken on lattices of size that can be studied in practice. In particular,
the emergent U($1$) symmetry of the VBS order parameter implies that the component not locked by the boundaries can survive in the form of significant
fluctuations up to very large system sizes, but this aspect of the ordering may be completely missed if one only examines the boundary-induced component 
of the order parameter. While this effect by itself would probably not lead to wrong conclusions regarding the presence or absence of VBS order, it is still
important for explaining results that would otherwise seem inconsistent with each other (e.g., when comparing the total squared order parameter and
a direct boundary induced order parameter, as was done here in Sec.~\ref{sec:jq2}). The results presented here suggest that the best quantity for extrapolating
the order parameter to infinite system size is the total (sum of the $x$ and $y$ components) long-distance correlation functions on $L\times L$ periodic lattices. 
Non-square lattices can lead to non-monotonic finite-size behavior.

Some of the small-system behaviors pointed out here are generically well known and not limited to VBS order. There are also many examples of finite-size scaling of results 
for small lattices leading to wrong conclusions of the nature of the ground state. For example, in Refs.~\onlinecite{parola93} and \onlinecite{ihle99} a spin liquid ground  
state was claimed to exist in a 2D system of weakly coupled $S=1/2$ Heisenberg chains. When QMC results for larger systems became available,\cite{sandvik99} they 
showed a cross-over of the scaling and an asymptotic behavior in accord with a N\'eel state for any value of the inter-chain coupling. 

The additional complications due to emergent U($1$) symmetry \cite{senthil04a,senthil04b,sandvik07} are more specific to VBS ordering. Open edges twist the vector 
order parameter $(D_x,D_y)$ in ways which depends on the model and the nature of the edge. For a VBS there is no ``neutral'' edge; any boundary affects the 
ordering pattern in its neighborhood. While in the bulk VBS, in the thermodynamic limit, only one of the components can survive in a columnar state, at edges they 
can both be present. Due to the large length-scale of the cross-over from the U($1$) symmetric order parameter, both components can also survive in the interior 
of large systems. It would be interesting to study this phenomenon also in systems with a more complicated (larger unit cell) VBS order parameter.

It should be noted that, although the concept of emergent U($1$) symmetry of VBSs was developed in the context of deconfined quantum-critical points and 
has been confirmed in the case of $J$-$Q$ models,\cite{sandvik07,jiang08,lou09} this aspect of VBS order is most likely very general and manifested also in systems 
that are not very close to such critical points (in some extended parameter space)---in 2D systems in which ``angular'' VBS fluctuations are possible once the 
correlation length is several lattice constants or larger. The U($1$) related boundary effects should be absent in cases where the angular fluctuations are absent, 
e.g., in the case of staggered VBS states.\cite{banerjee11a,sen10}

For the purpose of detecting VBS order, an important aspect of the critical scaling is that, once a critical point has been identified, upward deviations 
from the power-law behavior, as seen in Fig.~\ref{critical} at $J/Q_2=0$ and $0.03$ in the $J-Q_2$ model, demonstrate an ordered state although this may not be 
apparent when carrying out extrapolations of the order parameter in $1/L$ (as in Fig.~\ref{peri}). In general, in a model with some tunable parameter that can 
bring it into or out of a VBS state, it may be easier to detect a phase transition than to extract the exact value of the order parameter close to such a point. 
On the one hand, many frustrated systems may have VBS states that are always only weakly ordered and, hence, close to a quantum critical point (or weakly first-order 
transition) in some extended parameter space. Such systems should exhibit near-critical scaling on small lattices. On the other hand, if no critical scaling can 
be detected, and instead the order parameter correlation function decays exponentially fast with distance (or shows a tendency to decay faster than a power law), one 
can rather safely conclude that there is no VBS long-range order. Also with this approach, one can of course not expect to draw reliable conclusions unless 
the system sizes are sufficiently large (and how large that is depends on the model).

A striking behavior that may be particularly prominent in the case of VBS order was found here for lattices in the form of long cylinders, of size $L_x\times L_y$ 
with $L_x \to \infty$ and finite even $L_y$. In this geometry the order is unstable, and the system exhibits only short-range VBS correlations, until $L_y$ exceeds 
some threshold that can be very large (perhaps 3-4 times the VBS correlation length, according to results for the $Q_3$ model). Long cylinders are 
therefore not ideally suited for determining the nature of the 2D state in which VBS order is a possibility (systems with a small fixed $L_x/L_y$ normally being better). 
In particular, the method of positively confirming a $Z_2$ spin liquid by the absence of of order on even-$L_y$ systems is not applicable in the ``yes-no'' sense 
proposed in Ref.~\onlinecite{yao11}. Instead, the finite-size behavior has to be tracked as in any other extrapolation method. The correlation length as a function 
of even $L_y$ should converge for a spin liquid and diverge for a VBS, as in Fig.~\ref{cylsummary}, but it may not be easy in practice to determine which of these 
behaviors applies. 

\subsection{Comment on the possibility of a spin-liquid state in the $J_1$-$J_2$ Heisenberg model}

One motivation for the present study was to provide guidance on detecting VBS order---or, alternatively, showing the absence of such order---in calculations 
for frustrated 2D models. The lattice sizes reachable for such systems with unbiased calculations, primarily using the DMRG method,\cite{stoudenmire12,yan11,jiang11} 
are still very limited. Methods based on tensor-product states,\cite{murg09,wang11,evenbly10} beyond matrix-product states (which are closely related to the DMRG scheme), 
are still typically too much affected by various truncation errors and approximations to be considered completely unbiased. The following discussion will therefore be 
primarily aimed at DMRG calculations, although many of the issues would apply more generally.

The issues raised here have particular relevance in the context of a recent DMRG study of the $J_1$-$J_2$ Heisenberg model on the square lattice.\cite{jiang08,jiang11} 
Several different ways of analyzing VBS correlations were argued to consistently show the absence of VBS order and positively confirm the properties of a $Z_2$ spin 
liquid. However, many of the results presented can also be explained by a VBS state, at least in some part of the non-magnetic phase, according to the results 
obtained here. The key points supporting this view are summarized next.

In Fig.~3 of Ref.~\onlinecite{jiang11}, second-order polynomial fits to the VBS order parameter for $2L\times L$ cylinders with $L\le 10$ are shown. The fact that 
these fits extrapolate to negative values in the thermodynamic limit was taken as evidence for the absence of VBS order. However, this kind of behavior is also 
observed for the $Q_2$ model on small lattices, as seen in Figs.~\ref{peri} and \ref{open} of the present paper, even though the order parameter 
of this model is as large as $20\%$ of the maximum possible value. If VBS order exists also in the non-magnetic phase of the $J_1$-$J_2$ Heisenberg model, one should 
not expect it to be very strong. Therefore, the finite-size behavior seen in Fig.~3 of Ref.~\onlinecite{jiang11} is at least qualitatively what would be expected even if 
the state is a VBS. It should be noted that the fact that the fitted functions extrapolate to negative values is in itself a clear sign of the chosen functional forms  
not being correct, as the squared order parameter cannot be negative. Thus, there must necessarily be a cross-over to a different form for larger systems---either to a 
pure $1/L^2$ form, if there is no long-range order, or to an exponentially rapidly convergent form tending to a non-zero value. The results for small systems cannot
distinguish between these different asymptotics.

The finite-size extrapolation issues may clearly also affect the determination of the transition point between the N\'eel antiferromagnet and the non-magnetic state at 
$g=J_2/J_1\approx 0.4$ (while the transition point into the stripe antiferromagnet at $g=J_2/J_1\approx 0.6$ is much easier to extract due to it being clearly first order). 
The transition point $g\approx 0.41$ was determined in Ref.~\onlinecite{jiang11} 
based on extrapolations of the N\'eel order parameter $\langle M^2\rangle$ using second-order polynomials, and 
these should be affected by similar problems as those pointed out here for the VBS scaling (and it is also well known that polynomials higher than second order 
have to be used to extrapolate N\'eel order correctly based on small systems, even in the strongly order Heisenberg model \cite{sandvik99,sandvik10b}). The N\'eel 
order should therefore survive up to somewhat larger $g$ values. Thus, at $g=1/2$, on which most of the analysis of the VBS scaling was focused in 
Ref.~\onlinecite{jiang11}, the system may be rather close to the transition point. If VBS order exists in the nonmagnetic phase, it would therefore likely be very 
weak at this point. In Fig.~3(a) of Ref.~\onlinecite{jiang11}, the maximal value of the order parameter $\langle D_y^2\rangle$, at $g$ just below $0.6$, is close 
to the values for the $Q_2$ model in Fig.~\ref{open} of the present paper. Thus, if the $J_1$-$J_2$ model has VBS order, its peak value should be about 
$10-20\%$ of that of a perfect columnar state. It would be better to analyze the VBS correlations closer to the maximal value, where the extrapolation problems
are minimized. 

\begin{figure}
\includegraphics[width=7.5cm, clip]{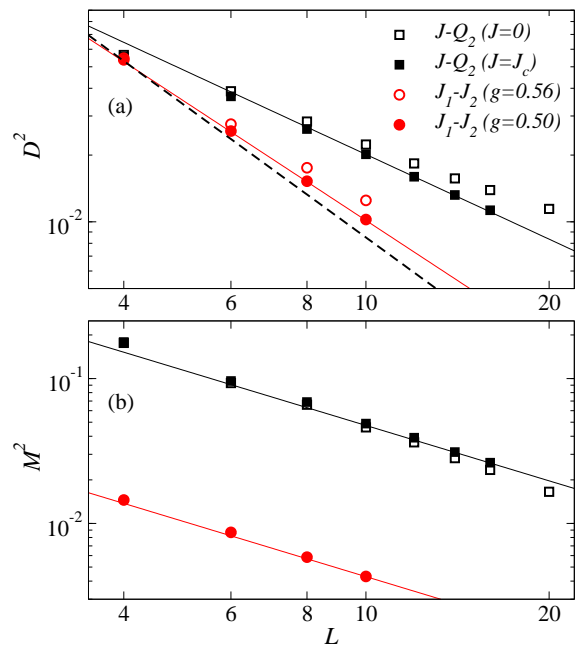}
\vskip-1mm
\caption{(Color online) Finite-size scaling of the squared VBS order parameter (a) and staggered magnetization (b) calculated on the 
central $L\times L$ square of cylinders of size $2L\times L$. DMRG Results for the $J_1$-$J_2$ model at $g=0.50$ and $0.56$, from Figs.~2(a) 
and 3(a) of Ref.~\onlinecite{jiang11}, are compared with QMC results for the $J$-$Q_2$ model at its critical point, $(J/Q_2)_c=0.0447$, and 
at $J=0$. In (a) the VBS $y$ component of the $J_1$-$J_2$ model and the $x$ component (the larger component) of the $J$-$Q_2$ are shown.
The line drawn close to the $J/Q_2=0.0447$ points has slope $-1.27$, corresponding to the critical exponent $\eta=0.27$ (as in Fig.~\ref{critical}), 
and that going through the $g=0.50$ points has slope $-1.8$. The dashed line has slope $-2$, corresponding to the expected asymptotic behavior in 
a non-VBS state. In (b) both lines have slope $-1.27$ ($\eta=0.27$).}
\label{open_c}
\vskip-1mm
\end{figure}

As discussed in Sec.~\ref{sec:critical}, in systems where there is a quantum phase transition into the state of interest, the best way to deduce the nature 
of that state may be to first carefully examine the phase transition. If there is critical scaling, deviations from the power-law form of the order parameter away 
from the critical point can be a good signal of long-range order. However, as seen in the scaling plot for the near-critical $J$-$Q_2$ model in 
Fig.~\ref{critical}, if the accessible system sizes are only up to $L\approx 10$, even a system in which the VBS order parameter is as large as $20\%$ 
of the maximum value may in practice not be distinguishable from a critical system when analyzing the order parameter fluctuations. If the non-magnetic
state of the $J_1$-$J_2$ Heisenberg model also has long range order, then one should expect a similar behavior. 

Re-plotting the $g=0.5$ and $0.56$ data for the VBS $y$ component of Fig.~3(a) of Ref.~\onlinecite{jiang11} on a log-log scale, one can indeed observe behaviors close to 
power laws, as shown in Fig.~\ref{open_c}(a). In the same graph data for the $J$-$Q_2$ model at $J=0$ and $J_c=0.0447$ are also graphed. In this case the $x$ component of the order 
parameter is shown, which, as seen in Fig.~\ref{open}, in this system is larger than the $y$ component and is the one surviving in the thermodynamic limit. In the $J_1$-$J_2$ model 
it is instead the $x$ component that is somewhat larger.\cite{privatecomm} 

The comparison of the two models is complicated by the fact that the average induced $x$ order was subtracted in the definition used in Ref.~\onlinecite{jiang11}. That 
induced order is very small, however,\cite{privatecomm} unlike what it is in the $Q_2$ model (which, may indicate that the VBS order, if it exists in the $J_1$-$J_2$ model, 
is $y$-oriented on the cylindrical $2L\times L$ systems, as was also noted in Ref.~\onlinecite{jiang11}).

For the open-edge $2L\times L$ cylinders used in Fig.~\ref{open_c}(a), the $J$-$Q_2$ results do not exhibit quite as good scaling as in the case of the periodic $L\times L$ 
systems in Fig.~\ref{critical}, but for large systems the behavior is still consistent with an exponent $\eta \approx 0.3$. The  $J_1$-$J_2$ results for $g=0.5$ follow a different 
behavior, however, decaying as $L^{-\alpha}$ with $\alpha \approx 1.8$. This is quite close to $\alpha=2$, which is expected deep inside a non-VBS phase. For $g=0.56$ 
the data for the larger sizes deviate significantly upward from the $g=0.5$ points and cannot be fitted very well to a power law. Tthe slope on the log-log scale is
$\approx -1.52$ for a line drawn through the $L=8$ and $10$ points, but the data for smaller systems fall above the fitted line, showing a flattening out with increasing size. 
The reduction of the rate of decay is opposite to the expectation for a spin liquid and an indication that the system is VBS ordered in the infinite-size limit. 

The behavior at $g=0.5$ is puzzling. Since the VBS order parameter here follows quite close to the form expected in a spin liquid, one may conclude that this 
is what it is, and the deviations from the $\sim 1/L^2$ form are due to remaining size effects (i.e., the system size is not yet much larger than the correlation length). 
A possibility suggested by the behavior observed in Fig.~\ref{open_c} is that the $J_1$-$J_2$ model has a spin liquid phase following the N\'eel phase above 
$g\approx 0.4$, followed in turn by a VBS at larger $g$ (since the $g=0.56$ results seem more indicative of weak VBS order). Another possibility is that there is no 
spin liquid, but the N\'eel--VBS transition takes place at $g$ significantly larger than previously believed, so that $g=0.5$ would actually still be inside the N\'eel 
phase. Looking at the raw data for the sublattice magnetization in Fig.~2(a) of Ref.~\onlinecite{jiang11}, it appears that this possibility cannot be ruled out (considering 
again also the fact that the second-order polynomial fits used should lead to an under-estimation of the critical $g$ where the N\'eel order vanishes). The behavior of the 
triplet gap in Fig.~2(b) seems to go against this scenario, however, although the way the gap was extracted, by targeting higher states obtained while keeping the edges 
in the ground state, may lead to strong corrections to the gap scaling. 

To investigate possible near-criticality in the N\'eel order parameter, the results from Fig.~2(a) of Ref.~\onlinecite{jiang11} for $\langle M^2\rangle$ at $g=0.5$ are re-plotted 
on a log-log scale in Fig.~\ref{open_c}(b). Interestingly, the behavior follows closely a power law, with an exponent $\eta$ very similar to that of the $J$-$Q_2$ model. This could 
indicate that the transition out of the N\'eel state indeed takes place close to $g=0.5$ and is in the same universality class as the $J$-$Q$ model. Note that, within the deconfined 
quantum criticality theory,\cite{senthil04a} this kind of criticality of the magnetic order would not necessarily require that the VBS order emerges at this point as well, because 
the exponents associated with the N\'eel order parameter are not affected by the VBS (since the operator causing the VBS order is dangerously invariant). Clearly there is not 
sufficient data here to make any firm conclusions about this scenario of a N\'eel to spin liquid transition, possibly followed by a subsequent liquid to VBS transition at
higher $g$, but the behavior is intriguing and deserves further tests. 

\begin{figure}
\includegraphics[width=7.5cm, clip]{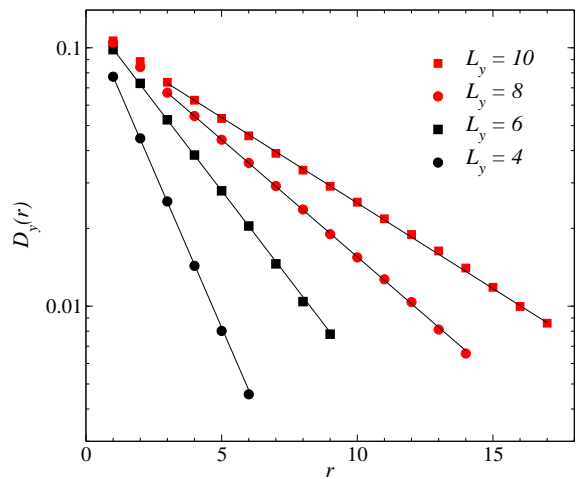}
\vskip-1mm
\caption{(Color online) VBS $y$ order parameter component induced by edges modified to break the $y$ translational symmetry (as explained in Fig.~\ref{mlatt})
in the $Q_2$ model. Here $r$ is defined as the distance from the second column of spins away from the edge, since the edge modification extends to this location. 
The lines show exponential fits, with decay lengths $1.8$ ($L_y=4$), $3.2$ ($L_y=6$), $4.8$ ($L_y=8$), and $6.6$ ($L_y=10$).}
\label{edgq2}
\vskip-1mm
\end{figure}

An important aspect of the analysis of Ref.~\onlinecite{jiang11}, cited as positive evidence for a $Z_2$ spin liquid, is the behavior of the order parameter on 
infinitely long cylinders. There is an even-odd effect that had previously been found in liquid states of quantum dimer models:\cite{yao11} For odd $L_y$ and even 
$L_x\to \infty$, an $x$-oriented order parameter $\propto {\rm exp}(-L_y/\xi_y)$ is induced because of geometric frustration effects. For even $L_y$ no order is 
observed at all, regardless of the type of VBS (horizontal or vertical columns) favored by the edges. Unfortunately, odd-$L_y$ $J$-$Q$ cylinders cannot be studied 
with the QMC method used here, because of sign problems. However, based on the results presented here for even $L_y$ it is already clear that this kind of test 
for a $Z_2$ spin liquid may not be that useful in practice, because VBS order does not exist on the infinitely long cylinders (for $L_y$ up to some critical width 
that can be expected to be inaccessible in practice for systems that are weakly to moderately ordered in the 2D limit). In Ref.~\onlinecite{jiang11} it was 
implicitly assumed that any system with 2D VBS order will exhibit such order also on long thin cylinders. 

It is also interesting to note that the induced order parameter as a function of the distance from a modified edge of systems in the $L_x \to \infty$
limit is very similar in the $J_1$-$J_2$ and $J$-$Q_2$ models. Fig.~\ref{edgq2} shows results for the pure $Q_2$ model on cylinders of width $4-10$ in which 
the edge has been modified to break the $y$ translational symmetry, as discussed in Sec.~\ref{sec:boundary} and illustrated in Fig.~\ref{mlatt}. Here
cylinders with aspect ratio $L_x/L_y=16$ were used (which is large enough to accurately represent the $L_x \to \infty$ limit). The edge-induced $x$ and $y$ 
order parameters both decay exponentially, with very similar decay lengths that are also close to the correlation lengths graphed in Fig.~\ref{cylsummary} 
(obtained from correlation functions on systems with all periodic boundaries). The $y$ decay lengths are always marginally larger. For $L=6$ and $8$, the decay 
lengths are about $1.5$ times those in the $J_1$-$J_2$ model at $g=0.5$, for which data were shown in Fig S6(b) of Ref.~\onlinecite{jiang11}. 

As discussed above and seen clearly in Fig.~\ref{open_c}, the VBS order parameter is likely significantly suppressed at $g=0.5$ relative to what 
it is close to its maximum in this model (which appears to be a bit above $0.56$). One can therefore expect to see decay lengths as large as those in the 
$Q_2$ model for larger $g$ (close to $0.6$). The rapid decay was in Ref.~\onlinecite{jiang11} interpreted as the system being insusceptible to VBS ordering even 
in the presence of, at first sight, very favorable conditions for inducing it. Again, when analyzed in light of the known physics of the $J$-$Q_2$ model, the 
results cannot be distinguished from those of a rather substantially ordered VBS. It would be illuminating to have $J_1$-$J_2$ data for $L_y>8$, to see if the
decay length continues to grow or saturates.

In Ref.~\onlinecite{jiang11} the size dependence of the entanglement entropy was also used as positive evidence of a $Z_2$ spin liquid. It would be very
interesting to compute this quantity also for the $J$-$Q$ models. It is clear that the non-trivial aspects of the VBS fluctuations could lead to behaviors
not predicted in the strong-VBS limit. Since the system on small lattices and cylinders resembles a spin liquid, it would not be surprising if the
corrections to the area law of the entanglement entropy are also similar, up to some large size where the true asymptotic VBS behavior sets in. QMC calculations of 
the entanglement entropy of the $J$-$Q$ models will be carried out in future studies, using the recent developments of methods to study the Renyi versions
of the entropies.\cite{hastings10,kallin11} This should clarify whether the constant deviation from the area law cited in Ref.~\onlinecite{jiang11} is 
really unique to $Z_2$ spin-liquids, or whether they can also appear (for lattices of practically reachable size) in weakly ordered VBS states. The scaling
of the entanglement entropy at a deconfined quantum-critical point is also of interest here.\cite{swingle11}

The conclusion reached from the above comparisons of results for the $J_1$-$J_2$ model and the $J$-$Q$ models is that they exhibit rather similar behaviors, and, 
therefore, a VBS ground state of the $J_1$-$J_2$ cannot be excluded. Some of the $J_1$-$J_2$ results may also be consistent with a $Z_2$ spin liquid at $g \approx 0.5$, 
but the point to note here is that most of the results presented so far do not favor that kind of state over a VBS state. In particular, the claimed positive signals 
for a $Z_2$ spin liquid are also seen in the confirmed VBS state of the $J$-$Q$ models. If anything, the very similar behaviors seen in the near-critical $Q_2$ model and the 
$J_1$-$J_2$ models should tilt the balance further in favor of VBS order for $g=J_2/J_1$ close to $0.6$. The behavior at $g=0.5$ is very intriguing and not consistent 
with a near-critical VBS of the same kind as in the $J$-$Q$ models. It would be very useful to analyze the VBS and magnetic correlations further in this case, 
preferrably on larger lattices. 

It would also be good to know in greater detail the effects of truncation (the number of states kept) in the DMRG calculations. The error $\approx 10^{-7}$ in 
Ref.~\onlinecite{jiang11} refers to the missing weight in the density matrix. One can expect the errors in the wave function to be approximately the square-root of 
this error,\cite{stoudenmire12,schollwock11} but exactly how much the VBS correlations are affected, especially for the largest systems, is not entirely clear.

\subsection{Remarks on other potential spin liquids}

The results presented here also are relevant to studies of the kagome Heisenberg model, for which DMRG studies also have indicated a spin liquid 
state.\cite{jiang08,yan11} A VBS is another candidate state,\cite{singh07,evenbly10} which is not easy to exclude if the ordering is weak (which should be 
expected, if this kind of order is present). Since the most likely VBS patterns in this case are much more complicated than the columnar state of the $J$-$Q$ 
models discussed here (with the most likely candidate states having 12- or 36-spin unit cells), it is not possible to relate results in the same close manner 
as done above in the case of the $J_1$-$J_2$ model. Nevertheless, the issues pointed out here should be considered also when analysing the kagome system, in 
particular on long cylinders. It would be very desirable to reach larger $L_x\times L_y$ lattices with the aspect ratio $L_x/L_y$ kept fixed, although this seems 
difficult at present. It would also be good to push calculations based on the multi-scale entanglement renormalization ansatz (MERA) \cite{evenbly10} to higher 
precision. Such a calculation had previously seemed to confirm the VBS with $36$-site cell proposed earlier based on other techniques,\cite{marston91,singh07} but 
the energy reached was not as low as that found with DMRG\cite{jiang08,yan11} and exact diagonalization.\cite{nakano11,lauchli11}

The analysis and arguments presented in this paper also suggest that it would be very useful to add to the nearest-neighbor Heisenberg exchange some term that favors 
one of the VBS states proposed previously, and to study the phase transition out of this ordered state. Longer-range couplings may work, but some interaction similar 
to the multi-spin $Q$ terms discussed here could be even better suited for inducing the desired type of VBS. 

Spin liquid states have recently also been claimed to exist in electronic Hubbard models and frustrated spin models on the honeycomb 
lattice.\cite{meng10,varney11,reuther11} For the Hubbard model, 2D lattices with up to hundreds of sites were used.\cite{meng10} The VBS correlations in
this case decay very rapidly with distance, and the system does not seem to exhibit the kind of problematic scaling issues pointed out in this 
paper. On the other hand, work on effective spin models constructed to capture the putative spin liquid state have not so far been 
conclusive.\cite{clark11,reuther11,albuquerque11,mezzacapo12} Also here it would be useful to extend the models in such a way that a VBS phase transition 
can be studied. The VBS should then be the one to which the ``bare'' honeycomb model is the most susceptible (which may in itself not be easy 
to determine in this case).

\begin{figure}
\includegraphics[width=8.4cm, clip]{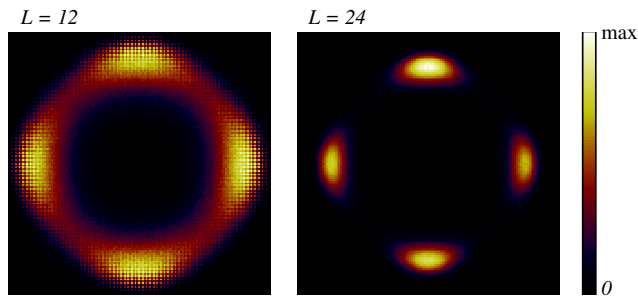}
\vskip-1mm
\caption{(Color online) VBS order parameter distribution $P(D_x,D_y)$ in the $Q_3$ model on periodic $L\times L$ lattices with $L=12$ (left)
and $L=24$ (right). The size of both squares corresponds to the full space of possible values of the components $D_x,D_y \in [-D_{\rm max},D_{\rm max}]$,
where $D_{\rm max}=3/8$ (for a perfect columnar VBS).}.
\label{histo3}
\vskip-1mm
\end{figure}

\subsection{Bench-mark challenge}

Finally, as a challenge to DMRG, tensor-product, and MERA techniques, it would be very interesting and useful to see these methods applied to $J$-$Q$ models as 
well. Comparing with the known phase diagram and critical behavior extracted on the basis of unbiased QMC simulations would be a very good test of the capabilities 
of these methods to capture non-trivial ground states and quantum phase transitions. If the outcome is positive, it may be very useful to systematically 
investigate the behavior when frustration is added to this model, as was recently done in an exact diagonalization study of a 2D model combining the $Q_2$ 
interaction with the frustrated $J_1$-$J_2$ Heisenberg model.\cite{nishiyama12}

\acknowledgments

I would like to thank Leon Balents, Ying-Jer Kao, Roger Melko, Rajiv Sing, Ying Tang, and Steve White for stimulating discussions and comments on the manuscript. 
I am also indebted to all the authors of Ref.~\onlinecite{jiang11} for providing numerical data from their manuscript and for discussing additional unpublished 
results. This research was supported by the NSF under Grant No.~DMR-1104708. Part of the work was carried out during a visit to National Taiwan University. 
I would like to thank its Center for Advanced Study in Theoretical Science for hospitality and support from Grant No.~NTU 10R80909-4.

\appendix

\section{U(1)--Z$_4$ cross-over of the VBS symmetry in periodic systems}
\label{appa}

The emergent U($1$) symmetry of a columnar VBS in the neighborhood of a critical point can be characterized by the probability distribution $P(D_x,D_y)$ generated
in QMC simulations on periodic $L\times L$ lattices. A systematic study aimed at extracting the scaling of the U($1$)-Z$_4$ cross-over length $\Lambda$ was presented 
in Ref.~\onlinecite{lou09}. Here additional results for the pure $Q_2$ and $Q_3$ models will be presented in order to facilitate comparisons with the boundary effects 
discussed in the main text. Specifically, it will be shown that the lack of $D_x$-$D_y$ symmetry on $2L \times L$ lattices, as seen in Fig.~\ref{perijq3} for the $Q_3$ 
model for all system sizes, is matched by a clear $Z_4$ symmetric order parameter on all $L\times L$ lattices. Conversely, the symmetry seen for the $Q_2$ model on large 
lattices in Fig.~\ref{perijq3} is consistent with only very small deviations (barely detectable) from U($1$) symmetry on $L\times L$ lattices with $L$ as large as $128$.

\begin{figure}
\includegraphics[width=7.5cm, clip]{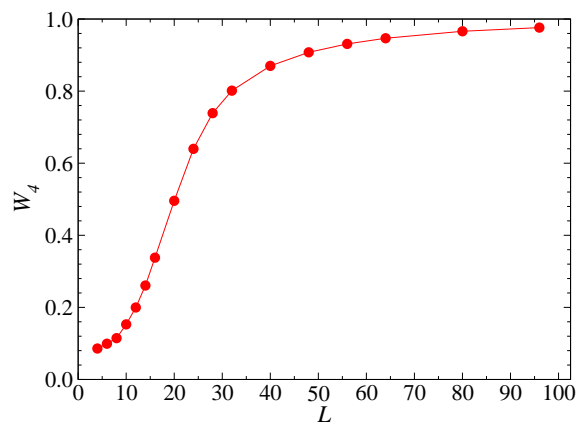}
\vskip-1mm
\caption{(Color online) Size dependence of the columnar anisotropy weight, defined in Eq.~(\ref{w4def}), of the VBS order parameter distribution in the $Q_3$ model.}
\label{w4}
\vskip-1mm
\end{figure}

In the projector QMC simulations, each generated configuration is associated with a pair of order parameters $(D_x,D_y)$, which are matrix elements of the
corresponding operators defined in Eqs.~(\ref{dxsum}) and (\ref{dysum}) computed in the valence bond basis. These matrix elements are of the form $3n/4N$, 
where $n$ is an integer in the range $[-N/2,N/2]$, with the extremal values corresponding to both the bra and ket state (making up the transition graph) 
having the same perfect columnar pattern of valence bonds of length one lattice constant. The histogram $P(D_x,D_y)$ is constructed based on these
matrix elements.

Fig.~\ref{histo3} shows results for the $Q_3$ model for $L=12$ and $L=24$. In this model the histogram $P(D_x,D_y)$ exhibits a distinct four-fold symmetry even for the 
smallest systems (also smaller than $L=12$, not shown here, where the discreteness of the distribution function also becomes apparent). The four peaks sharpen with 
increasing lattice size, and above some size the suppression of the weight between the peaks severely impedes QMC fluctuations between the peaks. In fig.~\ref{histo3}, 
the visibly different weight in the four peaks (with the right peak having the smallest weight) is a consequence of this rarity of ``instanton'' events between the peaks 
(i.e., the simulations ``get stuck'' in one quarter of the configuration space). It should be noted that this very slow simulation dynamics of the VBS order parameter 
does not affect the estimate of the total squared order parameter $\langle D^2\rangle$ and most other physical quantities of interest.

The degree of $Z_4$ symmetry of the order parameter can be quantified by the function
\begin{equation}
W_4 = \sum_{D_x}\sum_{D_y} P(D_x,D_y) \cos(4\phi_{xy}),
\label{w4def}
\end{equation}
where $\phi_{xy}$ is the angle corresponding to the point $(D_x,D_y)$. While this function (and the underlying probability distribution) is not a physical 
observable, in the sense that it is not a {\it bona fide} quantum mechanical expectation value, it nevertheless reflects the fluctuations of the VBS order 
parameter and can be used to characterize the the U($1$)-Z$_4$ cross-over.

\begin{figure}
\includegraphics[width=8.4cm, clip]{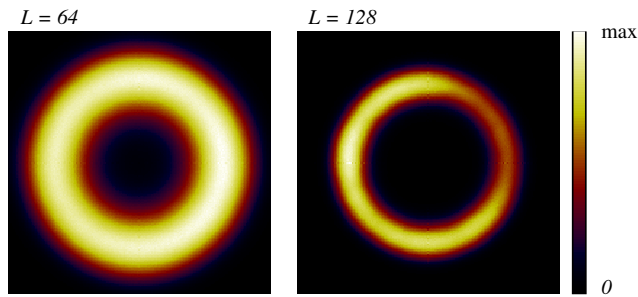}
\vskip-1mm
\caption{(Color online) VBS order parameter distribution $P(D_x,D_y)$ in the $Q_2$ model on periodic $L\times L$ lattices with $L=64$ (left)
and $L=128$ (right). The size of both squares corresponds to $10\%$ of the maximum value $D_{\rm max}/10$ of the components, $D_x,D_y \in [-D_{\rm max},D_{\rm max}]$, 
where $D_{\rm max}=3/8$ (for a perfect columnar VBS).}
\label{histo2}
\vskip-1mm
\end{figure}

\begin{figure}
\includegraphics[width=8.4cm, clip]{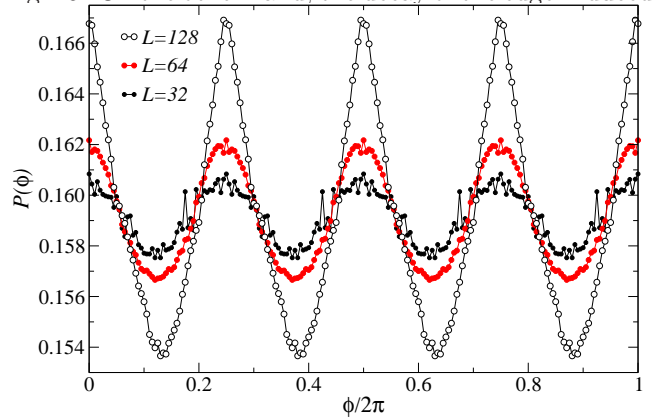}
\vskip-1mm
\caption{(Color online) Angular distribution of the VBS order parameter of the $Q_2$ model for system sizes $L=32$, $64$, and $128$. To improve the statistics,
these results were obtained by symmetrizing the distributions using the expected $90^\circ$ rotational symmetry. The jaggedness of the curves (especially for
$L=32$) is due to the discreteness of the allowed $(D_x,D_y)$ values (with $N$ possible values for each component).}
\label{phi}
\vskip-1mm
\end{figure}

Results as a function of $L$ for the $Q_3$ model are shown in Fig.~\ref{w4}. Here the convergence $W_4 \to 1$ when $L \to \infty$ is apparent, as would 
be expected for a columnar VBS in the thermodynamic limit. In principle the curve $W_4(L)$ could be used to define the length $\Lambda$, e.g., using 
$W_4(\Lambda)=1/2$, but there is clearly an arbitrariness in choosing the particular number. For studying the scaling of $\Lambda$ when some parameter 
of the Hamiltonian is changed (e.g., $J/Q_3$) this ambiguity does not matter. In Ref.~\onlinecite{lou09} curves $W_4(L)$ for different coupling rations were analyzed 
using standard finite-size scaling techniques, with the results that $\Lambda$ grows slightly faster than the correlation length; $\Lambda \sim \xi^{1+a}$ 
with $a\approx 0.2$.

Comparing with the behavior of the squared order parameters in Fig.~\ref{perijq3}, it can be noted that $\langle D_x^2\rangle$ approaches $0$ (and $\langle D_y^2\rangle$ 
tends to a non-zero value) very quickly above $L\approx 20$, which is approximately where $W_4(L)=1/2$ in Fig.~\ref{w4}. On the other hand, the decay of the
edge-induced $y$ component of the order parameter in Figs.~\ref{edg2q3} and ~\ref{xydecay} (where the system far from the edge has only $x$ order) gives a 
length $\approx 6.5$, which could also be taken as a practical definition of $\Lambda$. This length corresponds to $W_4 \approx 0.1$ in Fig.~\ref{w4}.

In contrast to the $Q_3$ model, in the $Q_2$ model no clear Z$_4$ symmetry is visible in $P(D_x,D_y)$ up to systems as large as $L=64$ and $128$,
as shown in Fig.~\ref{histo2}. These histograms are ring-shaped, although for $L=128$ the weight is not evenly distributed because of lack
of sufficient QMC statistics. The VBS angle fluctuates very slowly in simulations of large systems and very long runs are required in order to obtain symmetric
distributions. The data shown are based on $\approx 3.5 \times 10^8$ Monte Carlo sweeps for $L=64$ and $8\times 10^7$ for $L=128$ (which required more
than $10^4$ CPU hours in both cases). By symmetrizing the distributions using $90^\circ$ rotations, one can still detect small deviations from perfect
U($1$) symmetry, as shown in Fig.~\ref{phi}. The peak positions again correspond to a columnar state.

Note that in Fig.~\ref{histo2} the ring for $L=128$ is considerably thinner than for $L=64$, with the radius (the location of the maximum or average weight) 
remaining almost unchanged. This reflects an expected reduction of the fluctuations of the magnitude of the VBS order parameter with increasing system 
size.

Based on these results, the cross-over length-scale $\Lambda$ for the $Q_2$ model should be $\gg 128$, which explains why both order-parameter components 
are essentially equal for the largest systems in Fig.~\ref{peri_2}.

\end{document}